\documentclass{revtex4}

\usepackage{amsmath,amssymb}
\usepackage{color}
\usepackage{graphicx}
\usepackage{dcolumn}
\usepackage{bm}

\newcommand{\tr}{\operatorname{tr}}

\begin {document}

\title{Fluctuation Analysis of Time-Averaged Mean-Square Displacement for
Langevin Equation with Time-Dependent and Fluctuating Diffusivity}

\author{Takashi Uneyama}
\affiliation{%
  Faculty of Natural System, Institute of Science and Engineering,
  Kanazawa University, Kakuma, Kanazawa 920-1192,
  Japan
}%

\author{Tomoshige Miyaguchi}
\affiliation{%
Department of Mathematics Education, 
Naruto University of Education, Tokushima 772-8502, Japan}

\author{Takuma Akimoto}
\affiliation{%
  Department of Mechanical Engineering, Keio University, Yokohama 223-8522, Japan
}%


\date{\today}

\begin{abstract}
 The mean-square displacement (MSD) is widely utilized to study the dynamical
 properties of stochastic processes. The time-averaged MSD (TAMSD)
 provides some information on the dynamics which cannot be extracted from
 the ensemble-averaged MSD. In particular, the relative standard deviation
 (RSD) of the TAMSD can be utilized to study the long time relaxation behavior.
 In this work, we consider a class of Langevin equations which are
 multiplicatively coupled to time-dependent 
 and fluctuating diffusivities.
 Various interesting dynamics models
 such as entangled polymers and supercooled liquids can be
 interpreted as the Langevin equations with time-dependent and
 fluctuating diffusivities.
 We derive a general formula for the RSD of the TAMSD for
 the Langevin equation with the time-dependent and fluctuating
 diffusivity. We show that the RSD can be expressed in terms
 of the correlation function of the diffusivity. The RSD exhibits the crossover at
 the long time region.
 The crossover time is related to a weighted
 average relaxation time for the diffusivity. Thus the crossover time
 gives some information on the relaxation time of fluctuating diffusivity which
 cannot be extracted from the ensemble-averaged MSD.
 We discuss the universality and possible applications of the formula via some simple examples.
\end{abstract}

\maketitle


\section {Introduction}
\label{introduction}

The mean-square displacement (MSD) is one of the most commonly utilized
quantities to characterize the dynamical properties in experiments, theories, and
simulations.
Because a single-particle trajectory is a stochastic
variable, we need to perform averaging operations.
As the averaging operation, the ensemble average is widely employed.
The ensemble-averaged
MSD (EAMSD) is utilized, for example, to characterize the dynamical properties
of particles. In many systems,
the EAMSD shows a power-law type time dependence, i.e., the anomalous diffusion:
\begin{equation}
 \label{eamsd_definition}
  \langle [\bm{r}(\Delta) - \bm{r}(0)]^{2} \rangle \propto
  \Delta^{\alpha} .
\end{equation}
Here $\bm{r}(t)$ is a position of a particle at time $t$, $\Delta$ is the time difference, $\langle \dots \rangle$ represents the
ensemble average, and $\alpha > 0$ is the exponent which characterizes the
diffusion behavior ($\alpha < 1$, $\alpha = 1$, and $\alpha > 1$ correspond to
the subdiffusion, normal diffusion, and superdiffusion, respectively).
The anomalous behavior is observed in various systems ranging from a charge carrier transport in amorphous material\cite{Scher1975}, 
light diffusion\cite{Barthelemy2008}, polymeric materials\cite{Doi-Edwards-book},
to biological transports\cite{Caspi2000,Golding2006,Weigel2011,Tabei2013,Jeon-Tejedor-Burov-Barkai-SelhuberUnkel-BergSorensen-Oddershede-Metzler-2011,Jeon-Leijnse-Oddershede-Metzler-2013},
The diffusion behavior will depend on the time scale, and thus the exponent $\alpha$
may take several different values depending on $\Delta$.
For example, in
entangled polymers, the EAMSD of a segment exhibits four different regions which
reflect the crossovers between different characteristic relaxation time
scales\cite{Doi-Edwards-book}. In supercooled liquids, the EAMSD of a glass-forming particle strongly
depends on the temperature, and it shows a transient plateau. This
is considered as one evidence of the
cage-effect, which constrains the motion of the particle into a narrow
region\cite{Sciortino1996}. 

Although the EAMSD provides various useful information on the dynamical
properties, some properties cannot be extracted from the EAMSD. For example,
non-ergodic behavior cannot be analyzed from the EAMSD. For such a
purpose, the time-averaged MSD (TAMSD) can be utilized instead. The TAMSD is defined as
\begin{equation}
 \label{tamsd_definition}
  \overline{\delta^{2}(\Delta;t)} \equiv
  \frac{1}{t - \Delta} \int_{0}^{t - \Delta} dt' \,
  [\bm{r}(t' + \Delta) - \bm{r}(t')]^{2} ,
\end{equation}
where $\Delta$ and $t$ are the time difference and the observation time,
respectively.  If the system is ergodic and the time average is taken for a
sufficiently long observation time (at the limit of $t \to \infty$), the TAMSD
converges to the EAMSD with the equilibrium ensemble \cite{Akimoto2014}.
In molecular simulations and single-particle-tracking
experiments, it is not easy to calculate the
EAMSD. Instead, the TAMSD (or the average of the TAMSD over different
realizations and/or particles) is
widely used.  If the system is non-ergodic and/or the observation time
is not sufficiently long, the TAMSD does not coincide to the
EAMSD. In such a case, the TAMSD can be interpreted as a stochastic
variable. In some stochastic models of anomalous diffusion, such a randomness is intrinsic \cite{He-Burov-Metzler-Barkai-2008,Miyaguchi2011,Akimoto2013}.
In other words, TAMSDs remain random even when the observation time $t$ goes to infinity. Such an intrinsic randomness 
of the TAMSDs will be related to large fluctuations of the TAMSDs.
(The large fluctuations are actually observed in
single-particle-tracking experiments in living cells
\cite{Golding2006,Weigel2011,Tabei2013,Jeon-Tejedor-Burov-Barkai-SelhuberUnkel-BergSorensen-Oddershede-Metzler-2011,Jeon-Leijnse-Oddershede-Metzler-2013}.)
Thus it is important to calculate the statistical quantities such as the average
and standard deviation of the TAMSD.

The magnitude of the fluctuation of the
TAMSD can be quantitatively characterized by the relative fluctuation (RF) \cite{Akimoto2011,Uneyama-Akimoto-Miyaguchi-2012} or
the relative standard deviation (RSD) \cite{He-Burov-Metzler-Barkai-2008,Miyaguchi2011,Miyaguchi2011a}:
\begin{align}
 & \label{relative_fluctuation_definition}
 R(t;\Delta) \equiv \frac{\langle | \overline{\delta^{2}(\Delta;t)} -
 \langle \overline{\delta^{2}(\Delta;t)} \rangle | \rangle}
 {\langle \overline{\delta^{2}(\Delta;t)} \rangle} , \\
 & \label{relative_standard_deviation_definition}
 \Sigma(t;\Delta) \equiv \frac{\sqrt{\langle [\overline{\delta^{2}(\Delta;t)} -
 \langle \overline{\delta^{2}(\Delta;t)} \rangle]^{2} \rangle}}
 {\langle \overline{\delta^{2}(\Delta;t)} \rangle} .
\end{align}
The RF and RSD behave in a similar way, and it is reported that these quantities
can characterize some dynamical properties of the system
\cite{Akimoto2011,Miyaguchi2011a,Uneyama-Akimoto-Miyaguchi-2012}.
(If the second moment of $\overline{\delta^{2}(\Delta;t)}$ diverges, the
RSD diverges and the RF should be utilized to characterize the fluctuation of the TAMSD
\cite{Akimoto2014}.
In some literature, the squared RSD is utilized as the ergodicity
breaking parameter
\cite{He-Burov-Metzler-Barkai-2008,Burov-Jeon-Metzler-Barkai-2011,Cherstvy-Chechkin-Metzler-2013,Deng-Barkai-2009}.)
The RF and RSD analyses for the TAMSD are useful if the
systems are non-ergodic.
The $t$-dependence of the RF or RSD can be related to the ergodic
property of the system. For example, Deng and Barkai \cite{Deng-Barkai-2009}
analyzed the RSD of the TAMSD for the fractional Langevin equation and the
fractional Brownian motion. They obtained the analytic expression for
the RSD, and showed that the behavior of the RSD depends on the Hurst
parameter in a non-trivial way.

The RF and RSD analyses are also useful to study ergodic systems.
In the recent work\cite{Uneyama-Akimoto-Miyaguchi-2012}, the authors applied the RF analysis to the center
of mass motion in entangled polymer systems\cite{Doi-Edwards-book}. In entangled polymer systems,
the RF of the TAMSD shows the crossover 
behavior:
\begin{equation}
 R(t;\Delta) \propto
  \begin{cases}
   t^{-\beta} & (t \lesssim \tau_{c}') , \\
   t^{-0.5} & (t \gtrsim \tau_{c}') .
  \end{cases}
\end{equation}
Here $\beta < 0.5$ is the constant and $\tau_{c}'$ is the characteristic crossover
time. The crossover time $\tau_{c}'$ behaves in the same way as the longest
relaxation time (the disengagement time) $\tau_{d}$.
This means that the crossover time $\tau_{c}'$ characterizes the long
time relaxation in entangled polymer systems. 
(It would be natural to expect that the RSD of
the TAMSD also shows the similar crossover behavior, although the data are not
shown in the previous work.)
Interestingly, the EAMSD does not show such a crossover
around the longest relaxation time, and thus we consider that the TAMSD
is actually useful for the analysis of the long time relaxation behavior
in ergodic systems.
However, the reason why
$\tau_{c}'$ characterizes the long time relaxation behavior has not been theoretically
clarified yet.

One possible explanation is that the crossover originates
from the coupling between the dynamic equation for the
center of mass and the end-to-end vector \cite{Doi-Edwards-book}.
In the reptation model, a tagged polymer chain is
modeled as a polymer chain confined in a tube-like obstacle.
The polymer chain is allowed to move only along the tube. Due to this
constraint, the dynamic equation and the relaxation behavior become nontrivial.
(The reptation model can qualitatively reproduces the characteristic
dynamical properties, such as the relaxation modulus.)
The dynamic equation for the center of mass of the chain
can be explicitly expressed as \cite{Doi-Edwards-1978}
\begin{equation}
 \label{langevin_equation_reptation_model}
 \frac{d\bm{r}_{\text{CM}}(t)}{dt} = \sqrt{\frac{6 D_{\text{CM}}}{\langle
  \bm{p}^{2} \rangle}} \bm{p}(t) w(t) .
\end{equation}
Here $\bm{r}_{\text{CM}}(t)$ and $\bm{p}(t)$ are the center of mass position and
the end-to-end vector of an entangled polymer chain, respectively,
$D_{\text{CM}}$ is the diffusion coefficient for the center of mass, and $w(t)$ is the
one-dimensional Gaussian white noise. The first and second moments
of $w(t)$ are given as
\begin{equation}
 \label{first_and_second_moments_reptation_model}
 \langle w(t) \rangle = 0, \qquad
  \langle w(t) w(t') \rangle = \delta(t - t') .
\end{equation}
One important property of Eq.~\eqref{langevin_equation_reptation_model} is that
the noise $w(t)$ is multiplicatively coupled to another stochastic variable
$\bm{p}(t)$. Due to this multiplicative coupling, the magnitude of the random
motion of $\bm{r}_{\text{CM}}(t)$ directly depends on $\bm{p}(t)$. 
Although random variables $\bm{r}_\text{CM}(t)$ and $\bm{p}(t)$ 
are not statistically independent of each other, the coupling between
them is expected to be rather weak. (This is because the
dynamics of the end-to-end vector strongly depends on the resampling of new
segments at chain ends, and this resampling process is not directly
coupled to the dynamics of the center of mass.)
If we simply assume that they are statistically independent random variables (the decoupling approximation), we can interpret
Eq.~\eqref{langevin_equation_reptation_model} as the Langevin equation with
a time-dependent and fluctuating diffusivity.
Naively, we expect that such a multiplicative coupling causes the nontrivial
crossover behavior of the RF.

Similar time-dependent and fluctuating diffusivity has been reported for other systems. For
example, the diffusion of molecules in supercooled liquids is known to be
heterogeneous \cite{Sillescu-1999,Gotze-Sjogren-1992,Richert-2002,Berthier-Biroloi-2011}. This ``dynamic heterogeneity'' can be modeled
 by employing
time-dependent fluctuating diffusion coefficient. The simplest model may be the
two-state model \cite{Sillescu-1999} in which a tagged particle takes the slow state or fast state, and
the diffusion coefficients of the slow and fast states are different.
The intermittent search
strategies \cite{Benichou-Loverdo-Moreau-Voituriez-2011} also consist of
fast and slow diffusion modes. They are
considered to be important for rapid detection of
targets in biological systems such as foraging behavior of animals and
reaction pathways of DNA-binding proteins to the binding sites
\cite{Benichou-Loverdo-Moreau-Voituriez-2011}.
These models can be also interpreted as the Langevin equations with
time-dependent and fluctuating diffusivities.

Therefore, the analysis for a class of Langevin equations
with time-dependent and fluctuating diffusivity will provide useful
information for several different systems.
From the RF analysis result for entangled polymers,
the RF and RSD of the TAMSD are expected to be especially useful to quantify the
dynamical behavior. However, as far as the authors know,
theoretical analyses of the TAMSD for systems with time-dependent and fluctuating
diffusivities have not been reported.
In this work, we first introduce
a class of Langevin equations with time-dependent and fluctuating
diffusivity. Such a class of Langevin equations has not been studied in
detail. Then we analyze the RSD of the TAMSD, and derive a general formula
for the RSD.
We show that the RSD can be related to the time correlation
function of the diffusivity.
Our formula gives the relation between the crossover time
of the RSD and the relaxation time of the 
diffusivity. The crossover time is expressed in terms of a weighted average
relaxation time of the diffusivity.
We show the universality of our formula through some analytically solvable examples;
the (pure) reptation model for entangled polymers, and the
two-state model for the supercooled liquid with the Markovian and
non-Markovian transition dynamics.
Finally, we compare our analysis method with other analysis
methods, and discuss the properties of our method.
We also discuss the connection between the time-dependent and
fluctuating diffusivity and other models.

\section {Model}
\label{model}

In this work, we consider a class of Langevin equations with 
a time-dependent and fluctuating diffusivity.
As we mentioned, both the reptation model and the two-state model can be
interpreted as such Langevin equations.
In the reptation model \cite{Doi-Edwards-1978}, the one-dimensional thermal noise is multiplicatively coupled
to the three-dimensional end-to-end vector. On the other hand, in the two-state model,
the three-dimensional thermal noise is multiplicatively coupled to the scalar diffusion
coefficient. Although they are not equivalent, we may interpret these
models are special cases of a more general Langevin equation.

We consider a general multiplicatively coupled Langevin
equation model in an $n$-dimensional space \cite{Gardiner-book}.
For simplicity, we assume that no external force is applied.
The dynamic equation can be expressed as
\begin{equation}
 \label{langevin_equation}
 \frac{d\bm{r}(t)}{dt} = \sqrt{2} \bm{B}(t) \cdot \bm{w}(t) .
\end{equation}
Here, $\bm{r}(t)$ is the position, $\bm{B}(t)$ is the noise coefficient matrix,
and $\bm{w}(t)$ is the Gaussian white thermal noise.
The first and second moments of $\bm{w}(t)$ are
\begin{equation}
 \label{first_and_second_moments_w}
 \langle \bm{w}(t) \rangle = 0, \qquad
  \langle \bm{w}(t) \bm{w}(t') \rangle = \delta(t - t') \bm{1} ,
\end{equation}
where $\langle \dots \rangle$ represents the ensemble average and $\bm{1}$ is
the $n$-dimensional unit tensor.  We assume that $\bm{B}(t)$ obeys a stochastic
process which is stationary and independent of $\bm{r}(t)$ and $\bm{w}(t)$. Therefore, two
independent stochastic processes ($\bm{B}(t)$ and $\bm{w}(t)$) are
multiplicatively coupled in Eq.~\eqref{langevin_equation}.
As we show below, our model does not exhibit the anomalous
diffusion process, since $\bm{B}(t)$ obeys a stationary
stochastic process.
(A non-stationary process of $\bm{B}(t)$, such as the process with explicit time
dependence, generates anomalous diffusion
\cite{Fulinski-2011,Fulinski-2013,Massignan-Manzo-TorrenoPina-GarciaParajo-Lewenstein-Lapeyre-2014}.
In the followings, we
consider only stationary processes.)

The dynamics model for the noise coefficient matrix can be any stochastic processes,
such as the Langevin equation and the jump dynamics. The details
are not required for the analysis in the next section.
We need only several ensemble-averaged correlation functions.
For convenience, we define the instantaneous diffusion coefficient matrix
$\bm{D}(t)$ as
\begin{equation}
 \label{instantaneous_diffusion_coefficient_matrix}
 \bm{D}(t) \equiv  \bm{B}(t) \cdot \bm{B}^{\mathrm{T}}(t) .
\end{equation}
Conversely, we may interpret Eq.~\eqref{instantaneous_diffusion_coefficient_matrix} as the definition of the
noise coefficient matrix. That is, if we have the stochastic process for the
instantaneous diffusion coefficient $\bm{D}(t)$, the noise coefficient matrix can be defined
as the matrix square root (such as the Cholesky decomposition). The
instantaneous diffusion coefficient matrix $\bm{D}(t)$ should be positive
definite, and this condition guarantees the existence of the matrix square root.

The EAMSD is simply calculated to be
\begin{equation}
 \label{eamsd}
 \begin{split}
  \langle [\bm{r}(\Delta) - \bm{r}(0)]^{2} \rangle
  & = 2 \int_{0}^{\Delta} ds \int_{0}^{\Delta} ds' \,
  \langle \bm{B}(s) \cdot \bm{B}^{\mathrm{T}}(s') \rangle : \langle \bm{w}(s) \bm{w}(s') \rangle \\
  & = 2 \tr \langle \bm{D} \rangle \Delta ,
 \end{split}
\end{equation}
where the symbol ``$:$'' means a double dot product of tensors, i.e.,
$\bm{X} : \bm{Y} \equiv \sum_{ij} X_{ij} Y_{ij}$ for
second rank tensors $\bm{X}$ and $\bm{Y}$. In the last line of
Eq.~\eqref{eamsd}, we utilized the fact that the ensemble-average of the instantaneous diffusion
matrix becomes time-independent due to the time-translational invariance: $\langle \bm{D}(t) \rangle = \langle
\bm{D} \rangle$.
(The ensemble average of the instantaneous diffusion coefficient is
independent of time $t$, due to the stationarity.)
If we assume $\langle \bm{D} \rangle$ to be isotropic, we can
simply express $\langle \bm{D} \rangle$ as
\begin{equation}
 \langle \bm{D} \rangle = D_{\text{eff}} \bm{1} ,
\end{equation}
with $D_{\text{eff}}$ being the effective diffusion coefficient. Then
Eq.~\eqref{eamsd} is rewritten as:
\begin{equation}
 \label{eamsd_modified}
  \left\langle [\bm{r}(\Delta) - \bm{r}(0)]^{2} \right\rangle = 2 n D_{\text{eff}}
  \Delta,
\end{equation}
where $n$ is the dimension of the system.

The multiplicatively-coupled Langevin equation shown above cannot be expressed
as the generalized Langevin equation (GLE) with the Gaussian noise. Fox
\cite{Fox-1977} showed
that a GLE with the Gaussian noise can be characterized only by its memory
kernel. Therefore, if one obtains the EAMSD, the corresponding GLE is uniquely
determined. Because our model gives only the normal diffusion behavior, the
corresponding GLE would become a normal Langevin equation with
a Gaussian white noise which has no memory effect.  This apparent inconsistency
comes from the assumption that the noise is Gaussian. (As shown in the next
section, the fourth order moment of the noise behaves in a qualitatively
different way from the Gaussian noise.)
The simple dynamics
models such as the reptation model and the two-state model cannot be expressed
as the GLE with the Gaussian noise.

If the force is applied, we need to add the term proportional to the force
$\bm{F}(t)$ to the Langevin equation. Then Eq.~\eqref{langevin_equation} is
modified as
\begin{equation}
 \label{langevin_equation_with_force}
 \frac{d\bm{r}(t)}{dt} = \bm{\Lambda}(t) \cdot \bm{F}(t) + \sqrt{2}
 \bm{B}(t) \cdot \bm{w}(t) ,
\end{equation}
where $\bm{\Lambda}(t)$ is the time-dependent instantaneous mobility matrix. If we assume that
the fluctuation-dissipation relation of the second kind holds for the
instantaneous mobility, we have
\begin{equation}
 \label{instanteneous_mobility_matrix}
 \bm{\Lambda}(t) = \frac{1}{k_{B} T} \bm{D}(t) 
  = \frac{1}{k_{B} T} \bm{B}(t) \cdot
  \bm{B}^{\mathrm{T}}(t)  .
\end{equation}
with $k_{B}$ and $T$ being the Boltzmann constant and the absolute
temperature, respectively. Eqs.~\eqref{langevin_equation_with_force} and
\eqref{instanteneous_mobility_matrix} will be useful to study a particle
trapped in a potential or driven by an external force.

Before we proceed to the detailed analysis, we show that our general model
reduces to the reptaion and two-state models for some special cases. 
For a case where $n = 3$ and the noise coefficient matrix is given as
\begin{equation}
 \bm{B}(t) = \sqrt{\frac{3 D_{\text{CM}}}{\langle \bm{p}^{2} \rangle}}
 \frac{\bm{p}(t) \bm{p}(t)}{|\bm{p}(t)|},
\end{equation}
Eq.~\eqref{langevin_equation} reduces to the reptation model.
By introducing the one-dimensional Gaussian white noise $w'(t)$ as
\begin{equation}
  w'(t) \equiv \frac{\bm{p}(t)}{|\bm{p}(t)|} \cdot \bm{w}(t),
\end{equation}
Eq.~\eqref{langevin_equation} can be rewritten as follows:
\begin{equation}
 \label{langevin_equation_special1}
 \frac{d\bm{r}(t)}{dt} =
  \sqrt{\frac{6 D_{\text{CM}}}{\langle \bm{p}^{2} \rangle}} 
   \bm{p}(t) w'(t) .
\end{equation}
The first and second order moments of $w'(t)$ become
\begin{equation}
 \label{first_and_second_moments_special1}
 \langle w'(t) \rangle = 0, \qquad
  \langle w'(t) w'(t') \rangle
  = \delta(t - t') .
\end{equation}
Eqs.~\eqref{langevin_equation_special1} and \eqref{first_and_second_moments_special1} are
equivalent to the reptation model (Eqs.~\eqref{langevin_equation_reptation_model} and \eqref{first_and_second_moments_reptation_model}).
For the case where $n = 3$ and the noise coefficient matrix is
isotropic as
\begin{equation}
 \bm{B}(t) = \sqrt{2 D(t)} \bm{1},
\end{equation}
Eq.~\eqref{langevin_equation} simply reduces as follows:
\begin{equation}
 {d\bm{r}(t)}/{dt} = \sqrt{2 D(t)} \bm{w}(t) .
\end{equation}
This can be interpreted as the two-state model for supercooled
liquids or the trap model, if it is combined with appropriate transition dynamics for $D(t)$.

\section {Theory}
\label{theory}

The EAMSD cannot extract the information on the instantaneous
diffusion coefficient. As we mentioned, the fluctuation analysis of the TAMSD is
useful to characterize the long time relaxation behavior of entangled
polymers. In the reptation model,
the end-to-end vector is multiplicatively coupled to the
thermal noise in the Langevin equation. Naively,
the fluctuation of the TAMSD is expected to be governed by the dynamics
of the end-to-end vector. In the general Langevin equation model with
time-dependent diffusivity, the fluctuation of the TAMSD can be related to the
relaxation behavior of the noise coefficient matrix or the instantaneous
diffusion coefficient matrix. In this section, we analyze the RSD of the TAMSD
and derive a formula which relates the RSD and the time correlation functions of
$\bm{D}(t)$.

Because we are considering a stationary process for $\bm{B}(t)$, the ensemble average
can be evaluated rather straightforwardly.
By taking an ensemble average in Eq.~\eqref{tamsd_definition}, we have
\begin{equation}
  \langle \overline{\delta^{2}(\Delta;t)} \rangle
  =
  \langle [\bm{r}(\Delta) - \bm{r}(0)]^{2} \rangle
  =
  2 \tr \left\langle \bm{D} \right\rangle \Delta ,
\end{equation}
where the time-translational invariance $\left\langle [\bm{r}(t'+\Delta) - \bm{r}(t')]^{2}
\right\rangle = \left\langle [\bm{r}(\Delta) - \bm{r}(0)]^{2} \right\rangle$ and
Eq.~\eqref{eamsd} have been utilized. Then, the RSD of the TAMSD
(Eq.~\eqref{relative_standard_deviation_definition}) for the Langevin
equation~\eqref{langevin_equation} is given by
\begin{equation}
 \label{rsd_definition}
 \Sigma(t;\Delta) 
 =  \sqrt{
 \frac{\langle [\overline{\delta^{2}(\Delta;t)}]^{2} \rangle}
 {4 (\tr\left\langle \bm{D} \right\rangle )^2 \Delta^{2}}
 - 1 } .
\end{equation}
  
We need the explicit expression of $\langle [\overline{
  \delta^{2}(\Delta;t)}]^{2} \rangle $ to calculate the RSD.
 We can obtain a rather simple expression for $\langle [\overline{
  \delta^{2}(\Delta;t)}]^{2} \rangle $, although the detailed
  calculations become lengthy. After straightforward but long
  calculations, we have
\begin{equation}
 \label{tamsd_square_average_exact}
  \begin{split}
   \langle \overline{ [\delta^{2}(\Delta;t)}]^{2} \rangle
   & = \frac{8}{(t - \Delta)^{2}} \int_{0}^{t - \Delta} dt' 
   \int_{0}^{t'} dt'' \int_{t'}^{t' + \Delta} ds 
     \int_{t''}^{t'' + \Delta} ds' \, \langle \tr \bm{D}(s) \tr \bm{D}(s')
   \rangle \\
   & \qquad + \frac{32}{(t - \Delta)^{2}} \int_{0}^{t - \Delta} dt' 
   \int_{\max(0,t' - \Delta)}^{t'} dt'' 
   \int_{t'}^{t'' + \Delta} ds 
   \int_{t'}^{s} ds' \,
   \tr \langle \bm{D}(s) \cdot \bm{D}(s') \rangle .
  \end{split}
\end{equation}
See Appendix \ref{detailed_calculations_for_ensemble_average_of_squared_time-averaged_mean_square_displacement}
for detailed calculations. We consider the properties of two correlation functions in
Eq.~\eqref{tamsd_square_average_exact}, $\langle \tr \bm{D}(t) \tr
\bm{D}(t')  \rangle$ and $\tr \langle \bm{D}(t) \cdot \bm{D}(t')
\rangle$. We assume that the stochastic process $\bm{B}(t)$ 
is ergodic, and thus at the limit of  $|t - t'| \to \infty$, these
correlation functions can be decoupled:
\begin{align}
 & \label{correlation_fucntion_1_asymptotic}
 \langle \tr \bm{D}(t) \tr \bm{D}(t') \rangle \to
 (\tr \langle \bm{D} \rangle)^2 , 
 \\
 & \label{correlation_fucntion_2_asymptotic}
 \tr \langle \bm{D}(t) \cdot \bm{D}(t') \rangle \to
 \tr \left(\langle \bm{D} \rangle \cdot \langle \bm{D} \rangle\right) .
\end{align}
It would be convenient to rewrite two correlation functions by using Eqs.~\eqref{correlation_fucntion_1_asymptotic} and
\eqref{correlation_fucntion_2_asymptotic}, as follows:
\begin{align}
 & \label{correlation_fucntion_psi1_definition}
 \langle \tr \bm{D}(t) \tr \bm{D}(t') \rangle \equiv
 \left(\tr \langle \bm{D} \rangle\right)^2 
 \left[1 + \psi_{1}(t - t')\right] , \\
 & \label{correlation_fucntion_psi2_definition}
 \tr \langle \bm{D}(t) \cdot \bm{D}(t') \rangle \equiv
 n\tr \left(\langle \bm{D} \rangle \cdot \langle \bm{D} \rangle\right)
 \left[ \frac{1}{n} + \psi_{2}(t - t') \right] ,
\end{align}
where $\psi_{1}(t)$ and $\psi_{2}(t)$ represent four-body two-time correlation
functions. Both  $\psi_{1}(t)$ and $\psi_{2}(t)$ are symmetric in $t$
and approach to zero at $|t| \to \infty$.
(For one-dimensional systems ($n = 1$), $\psi_{1}(t) =
\psi_{2}(t)$. For $n \ge 2$, generally $\psi_{1}(t)$ and $\psi_{2}(t)$
do not coincide.)

By combining Eqs.~\eqref{rsd_definition},
\eqref{tamsd_square_average_exact},
\eqref{correlation_fucntion_psi1_definition},
and \eqref{correlation_fucntion_psi2_definition}, the squared RSD is
expressed as
\begin{equation}
 \label{rsd_square}
 \begin{split}
  \Sigma^{2}(t;\Delta)
  & = \frac{2}{\Delta^{2} (t - \Delta)^{2}} \int_{0}^{t - \Delta} dt' 
   \int_{0}^{t'} dt'' \int_{t'}^{t' + \Delta} ds 
     \int_{t''}^{t'' + \Delta} ds' \, \psi_{1}(s - s') \\
   & \qquad + \frac{8C}{\Delta^{2} (t - \Delta)^{2}} \int_{0}^{t - \Delta} dt' 
   \int_{\max(0,t' - \Delta)}^{t'} dt'' 
   \int_{t'}^{t'' + \Delta} ds 
   \int_{t'}^{s} ds' \,
   \left[ \frac{1}{n} + \psi_{2}(s - s') \right],
 \end{split}
\end{equation}
with $C$ being defined as
\begin{equation}
 \label{factor_c_definition}
C \equiv n \frac{\tr \left(\langle \bm{D} \rangle \cdot
  \langle \bm{D} \rangle\right)}{(\tr \langle \bm{D} \rangle)^{2}} .
\end{equation}
Note that if the
average diffusion coefficient matrix $\left\langle \bm{D} \right\rangle$ is
isotropic, we have $C = 1$.
In many practical cases, the observation time $t$ is much longer than
the time difference $\Delta$.
For such a case ($t \gg \Delta$),
Eq.~\eqref{rsd_square} is simplified as follows:
\begin{equation}
 \label{rsd_square_approximated2}
 \begin{split}
  \Sigma^{2}(t;\Delta)
  & \approx  \frac{2}{\Delta^{2} t^{2}} \int_{0}^{t} ds'' \,
  (t - s'') \int_{0}^{\Delta} ds
     \int_{0}^{\Delta} ds' \, \psi_{1}(s - s' + s'') \\
   & \qquad + \frac{4C}{3 n} \frac{\Delta}{t}
  + \frac{8 C}{\Delta^{2} t}
   \int_{0}^{\Delta} ds
   \int_{0}^{s} ds' \, (s - s') \psi_{2}(s') .
 \end{split}
\end{equation}

Moreover, if the characteristic relaxation time of $\psi_{1}(t)$ and
$\psi_{2}(t)$, $\tau$, 
is much longer than $\Delta$ ($\tau \gg \Delta$), Eq.~\eqref{rsd_square_approximated2} can be
further approximated:
\begin{equation}
 \label{rsd_square_final}
  \Sigma^{2}(t;\Delta)
  \approx \frac{2}{t^{2}} \int_{0}^{t} ds \,
  (t - s) \psi_{1}(s) .
\end{equation}
Thus the squared RSD becomes approximately independent of $\psi_{2}(t)$.
If $\psi_1(t)$ decays sufficiently fast as $t$ increases (strictly
speaking, if $\psi_1(t)$ decays faster than $t^{-1}$), we have the following 
asymptotic forms:
\begin{equation}
 \label{rsd_square_asymptotic}
  \Sigma^{2}(t;\Delta)
  \approx
  \begin{cases}
   \displaystyle
   \psi_{1}(0) & (t \ll \tau) , \\
      \displaystyle
   \frac{2}{t}  \int_{0}^{\infty} ds \, \psi_{1}(s) & (t \gg \tau) .
  \end{cases}
\end{equation}
Eqs.~\eqref{rsd_square_final} and
\eqref{rsd_square_asymptotic} are the main result of this
section.
For the case of $t \gg \tau$, the RSD behaves as
$\Sigma(t;\Delta) \propto t^{-1/2}$, which corresponds to the Gaussian
fluctuation. From Eq.~\eqref{rsd_square_final}, we find that the
$t$-dependence of the RSD is essentially determined only by
$\psi_{1}(t)$.
Therefore the crossover time $\tau_{c}$ is
related only to $\psi_{1}$. From Eq.~\eqref{rsd_square_asymptotic}, the crossover time
$\tau_{c}$ is estimated as
\begin{equation}
 \label{tau_c_explicit}
 \tau_{c} \approx \frac{2}{\psi_{1}(0)} \int_{0}^{\infty} ds \,
  \psi_{1}(s) .
\end{equation}
For a single exponential type relaxation ($\psi_{1}(t) = \psi_{1}(0)
e^{-t / \tau}$), this crossover time becomes:
\begin{equation}
 \label{crossover_time_single_exponential_relaxation}
 \tau_{c} \approx 2 \tau .
\end{equation}
As expected, the crossover time is proportional to the relaxation time,
although they are different by the numerical factor $2$.
In general, the correlation function $\psi_{1}(t)$ cannot be expressed
as a single exponential form but a sum of multiple
exponential relaxation modes.
Even in such a case, a similar
relation between the relaxation time and the crossover time
holds. For such a case, the relaxation time $\tau$ in
Eq.~\eqref{crossover_time_single_exponential_relaxation} is replaced by the weighted
average relaxation time for multiple exponential relaxation modes (with the weights
proportional to the amplitude of modes). This result justifies the use of the crossover time as the
characteristic relaxation time for systems with time-dependent
diffusivities, as long as $\psi_{1}(t)$ reflects the characteristic
relaxation at the long time scale.
As shown in Appendix
\ref{relation_between_relative_fluctuation_and_relative_standard_deviation},
the RF behaves in a similar way to the RSD. Thus we consider
that the empirical relation between the crossover time and the longest
relaxation time in the entangled polymers in the previous work
\cite{Uneyama-Akimoto-Miyaguchi-2012} is theoretically supported by
this work.

Before we proceed to calculations for some analytically solvable
models, we briefly consider the behavior of
the RSD in the case where $\Delta / t$ is not sufficiently
small. In such a case, the second and third terms in the right hand side
of Eq.~\eqref{rsd_square_approximated2} is not always negligible.
As before, we approximate the integrand in the third term in the right
hand side of Eq.~\eqref{rsd_square_approximated2} by $\psi_{2}(0)$. Then
we have
\begin{equation}
 \label{rsd_square_approximated_small_t}
 \Sigma^{2}(t;\Delta) \approx \psi_{1}(0) + 
  \frac{4 C}{3} \left[ \frac{1}{n} +
   \psi_{2}(0) \right] \frac{\Delta}{t} .
\end{equation}
If $t / \Delta \lesssim 4 C [1 / n + \psi_{2}(0)] / 3 \psi_{1}(0)$, the
contribution of the second term in the right hand side of
Eq.~\eqref{rsd_square_approximated_small_t} becomes non-negligible.
Roughly speaking, this term gives the correction, which is proportional to
$(\Delta / t)^{1/2}$, to the RSD.
It should be noted here that the $\Delta$ dependence of this correction is
rather simple. We may utilize the RSD data with different values
of $\Delta$ to obtain the data at the limit of $\Delta \to 0$ by the
extrapolation. In what follows, we will neglect this correction term
for simplicity.

\section {Examples}
\label{examples}

In this section, we apply the general formula (Eqs.~\eqref{rsd_square_final} and
\eqref{rsd_square_asymptotic} together with
Eq.~\eqref{correlation_fucntion_psi1_definition}) obtained in the
previous section 
to some analytically solvable models. We show the explicit forms of the RSD and
discuss how we can relate the long time relaxation behavior of systems to time-dependent diffusivity from the
fluctuation of the TAMSD.

\subsection {Reptation Model for Entangled Polymer}
\label{reptation_model_for_entangled_polymer}

As a simple model of entangled polymers, we consider the (pure) reptation
model\cite{Doi-Edwards-book}.
In the reptation model, the motion of a tagged polymer chain is modeled as one
of a polymer chain in a tube-like obstacle.
The dynamic equation is expressed as a one-dimensional
Langevin equation, and various dynamical properties can be analytically
calculated. For example, we can calculate the
shear relaxation modulus, the end-to-end vector relaxation function, and
the EAMSD.

As we mentioned, the dynamic equation for the center of mass in the
reptation model is given as Eq.~\eqref{langevin_equation_reptation_model}.
The instantaneous diffusion matrix $\bm{D}(t)$ becomes
\begin{equation}
 \bm{D}(t) = 3 D_{\text{CM}} \frac{\bm{p}(t) \bm{p}(t)}{\langle
  \bm{p}^{2} \rangle} .
\end{equation}
The effective diffusion coefficient $D_{\text{eff}}$ simply coincides with $D_{\text{CM}}$:
\begin{equation}
 D_{\text{eff}} = \frac{1}{3} \tr \langle \bm{D} \rangle 
  = D_{\text{CM}} .
\end{equation}
Under the decoupling approximation, the reptation model reduces to the
Langevin equation with the time-dependent and flucuating diffusivity,
and the general formula can be utilized.
The correlation function $\psi_{1}(t)$
(Eq.~\eqref{correlation_fucntion_psi1_definition}) becomes
\begin{equation}
 \label{correlation_function_psi1_reptation}
 \psi_{1}(t) = \frac{\langle \bm{p}^{2}(t) \bm{p}^{2}(0)
  \rangle}{\langle \bm{p}^{2} \rangle^{2}} - 1 .
\end{equation}
We need to calculate $\langle \bm{p}^{2}(t) \bm{p}^{2}(0) \rangle$ to
obtain the explicit expression for the RSD of the TAMSD.
The four-body two-time correlation function $\psi_{1}(t)$ can be analytically
evaluated. After long but straightforward calculations, we have the
following form for $\psi_{1}(t)$:
\begin{equation}
 \label{correlation_function_psi1_reptation_final}
  \psi_{1}(t) 
  =  \frac{16}{3 \pi^{2}} \sum_{k : \text{odd}} \frac{1}{k^{2}}
  E_{2}(k^{2} t / \tau_{d})  .
\end{equation}
Here $E_{m}(z)$ is the (generalized) exponential integral of the $m$-th
order \cite{NIST-handbook}, and $\tau_{d}$ is the disengagement time
\cite{Doi-Edwards-book} which corresponds to the longest relaxation time
in the reptation model.
The detailed calculations are summarized in Appendix
\ref{detailed_calculations_for_reptation_model}.

The behavior of the RSD of the TAMSD in the reptation model can be
calculated from Eqs.~\eqref{rsd_square_final},
\eqref{rsd_square_asymptotic} and
\eqref{correlation_function_psi1_reptation_final}.
The asymptotic forms can be calculated as follows. At $t = 0$,
$\psi_{1}(t)$ simply becomes
\begin{equation}
 \label{psi_zero_reptation}
 \psi_{1}(0)
   = \frac{16}{3 \pi^{2}} \sum_{k : \text{odd}} \frac{1}{k^{2}}
 = \frac{2}{3} .
\end{equation}
Here we have used $E_{2}(0) = 1$.
The integral of $\psi(t)$ over $t$ is calculated as
\begin{equation}
 \label{psi_integral_reptation}
   \int_{0}^{\infty} dt \, \psi_{1}(t)
   = \sum_{k : \text{odd}} \frac{16 \tau_{d}}{3 \pi^{2} k^{4}} 
   \int_{0}^{\infty} dz \, E_{2}(z) 
   = \frac{\pi^{2} \tau_{d}}{36},
\end{equation}
where we have used the integral formula for the exponential integral \cite{NIST-handbook}:
\begin{equation}
 \int_{0}^{\infty} dz \, E_{2}(z) = E_{3}(0) = \frac{1}{2} .
\end{equation}
From Eqs.~\eqref{psi_zero_reptation} and \eqref{psi_integral_reptation}, we have the following asymptotic forms for the
RSD of the TAMSD:
\begin{equation}
 \label{rsd_asymptotic_reptation}
 \Sigma(t;\Delta) \approx
  \begin{cases}
   \displaystyle \sqrt{\frac{2}{3}} & (t \ll \tau_{d}) , \\
   \displaystyle \sqrt{\frac{\pi^{2} \tau_{d}}{18 t}} & (t \gg \tau_{d}) .
  \end{cases}
\end{equation}
The crossover time $\tau_{c}$ is then estimated as
\begin{equation}
 \label{tau_c_explicit_reptation}
 \tau_{c} = \frac{\pi^{2} \tau_{d}}{12} \approx 0.822 \tau_{d} .
\end{equation}
Thus we find that $\tau_{c}$ is actually proportional to $\tau_{d}$.
Moreover, $\tau_{c}$ is closer to $\tau_{d}$ than the case
of the single relaxation time. This result is consistent with our
previous simulation results for the reptation model \cite{Uneyama-Akimoto-Miyaguchi-2012}. (The crossover time
of the RF, $\tau_{c}'$, is almost the same as the disengagement time.)
{ The analytic results shown above are obtained under the
decoupling approximation, which we employed without any justifications.
The decoupling approximation can be justified for the calculation of the RSD
of the TAMSD, and thus Eqs.~\eqref{rsd_asymptotic_reptation} and \eqref{tau_c_explicit_reptation}
can be also justified. See Appendix \ref{decoupling_approximation_for_reptation_model}.}

Here it would be worth noting that the integral in
Eq.~\eqref{rsd_square_final} can be analytically evaluated (although the
obtained expression becomes complicated). After
straightforward calculations, we have the following explicit expression
for the squared RSD:
\begin{equation}
 \label{rsd_square_explicit_reptation}
  \Sigma^{2}(t;\Delta) 
  = \frac{\pi^{2} \tau_{d}}{18 t}
  - \frac{\pi^{4} \tau_{d}^{2}}{270 t^{2}}
  +  \frac{32 \tau_{d}^{2}}{3 \pi^{2} t^{2}}  \sum_{k : \text{odd}} \frac{1}{k^{6}} 
E_{4}(k^{2}t / \tau_{d}) .
\end{equation}
This reduces to two asymptotic forms shown in
Eq.~\eqref{rsd_asymptotic_reptation}, at $t \ll \tau_{d}$ and $t \gg \tau_{d}$.

\begin{figure}[h]
 \includegraphics[width=1.0\linewidth,clip]{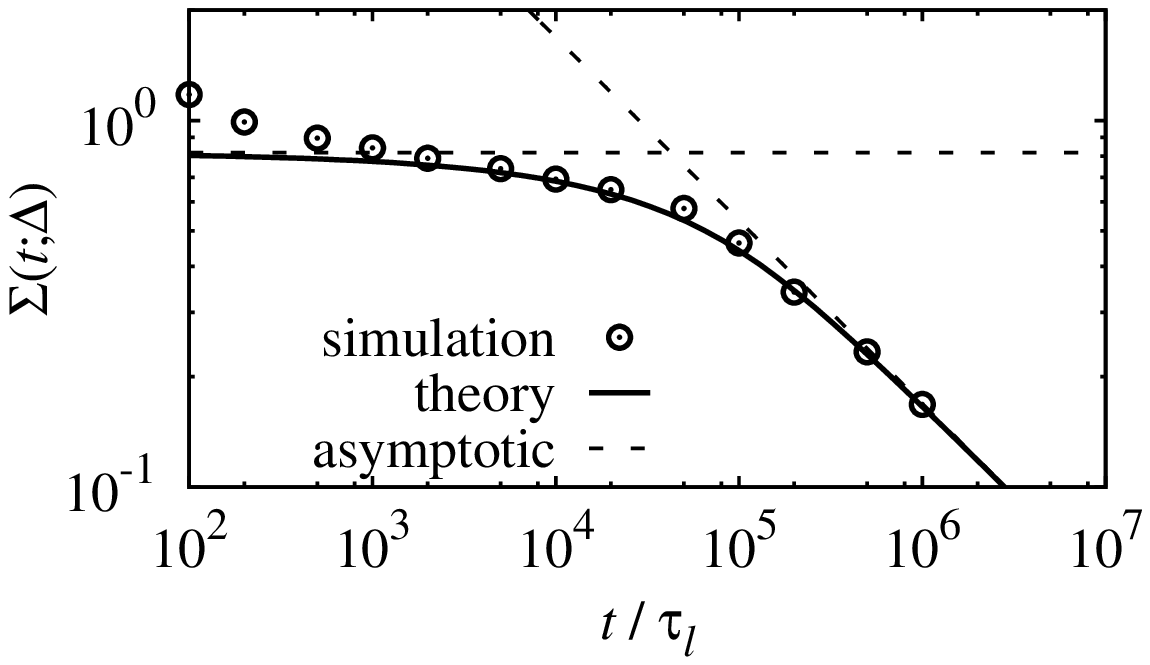}
\caption{\label{rsd_tamsd_reptation}}
\end{figure}

To validate our result, we perform a simulation for the discretized
version of the reptation model (the discrete reptation model) and
calculate the RSD of the TAMSD of the CM.
The dynamics of an entangled polymer is modeled by a
stochastic jump process. A polymer chain is expressed as a series of
discrete tube segments which have the constant size.
The chain randomly moves inside the tube (the reptation motion), and
the end segments are stochastically resampled.
(The details of the model and simulation method are described in the previous
work \cite{Uneyama-Akimoto-Miyaguchi-2012}.)
We show the RSD of the TAMSD for the number of tube segments per
chain $Z = 80$ in Figure \ref{rsd_tamsd_reptation}. The time difference
$\Delta$ is taken to be $\Delta = 10 \tau_{l}$ where $\tau_{l}$ is the
characteristic time scale for the longitudinal motion of a segment along
the tube.
We observe that our analytic expression (Eq.~\eqref{rsd_square_explicit_reptation}) and its
asymptotic forms (Eq.~\eqref{rsd_asymptotic_reptation}) are in good agreement with
the simulation result except for the small $t$ region.
This result supports the validity of our general formula 
(Eq.~\eqref{rsd_square_final}) and
its asymptotic forms (Eq.~\eqref{rsd_square_asymptotic}).

\subsection {Two-state Model for Supercooled Liquid}
\label{two_state_model_for_supercooled_liquid}

The dynamics of supercooled liquids have been extensively studied by
experiments, theories, and simulations
\cite{Sillescu-1999,Gotze-Sjogren-1992,Richert-2002,Berthier-Biroloi-2011}.
In the molecular dynamics (MD) simulations,
the motion of each particle can be observed and various
statistical quantities can be calculated. One important finding by the MD simulations is the
``dynamic heterogeneity \cite{Yamamoto-Onuki-1998,Yamamoto-Onuki-1998a}.''
The mobility or the diffusivity of a particle
strongly fluctuates spatially and temporally.  The dynamic heterogeneity is
considered as a characteristic property of supercooled or glassy liquids. Many
theoretical and experimental studies have been conducted to observe and characterize
the dynamic heterogeneity. The two-state model is a simple and analytically
solvable theoretical model which takes into account the dynamic
heterogeneity \cite{Sillescu-1999}.

In the two-state model, dynamics of a tagged particle is considered. The position of the
tagged particle at time $t$, $\bm{r}(t)$, obeys the following Langevin equation:
\begin{equation}
 \label{langevin_equation_two_state_model}
  \frac{d\bm{r}(t)}{dt} = \sqrt{2 D(t)} \bm{w}(t) .
\end{equation}
Here $D(t)$ is the time-dependent diffusion coefficient.
The particle has a state and the state is time-dependent.
We express the state of the particle at time $t$ as $h(t)$, and this
$h(t)$ can take either the fast ($f$) or slow ($s$) state.
The fast and slow states have different diffusion coefficients, and
thus the diffusion coefficient $D(t)$ is expressed as
\begin{equation}
 \label{diffusion_equation_two_state_model}
 D(t) =
  \begin{cases}
   D_{f} & (\text{for $h(t) = f$}) , \\
   D_{s} & (\text{for $h(t) = s$}) . 
  \end{cases}
\end{equation}
Here, $D_{f}$ and $D_{s}$ are diffusion coefficients of the fast and slow
states ($D_{f} > D_{s}$). We describe the probability that the particle
is at state $h$ at time $t$ as
$P_{h}(t)$. 
The four-body two-time correlation function $\psi_1(t)$ is given by
\begin{equation}
 \label{psi1_two_state}
 \psi_{1}(t) \equiv \frac{\langle D(t) D(0) \rangle}{\langle D
  \rangle^{2}} - 1.
\end{equation}
We express equilibrium fraction (equilibrium probability) of the state
$h$ as $\phi_{h}$ ($\phi_{h} \equiv \langle P_{h} \rangle$).
Then the effective diffusion coefficient can be expressed as
\begin{equation}
 \label{effective_diffusion_coefficient_two_state_model}
 D_{\text{eff}} = \langle D \rangle = D_{f} \phi_{f} + D_{s} \phi_{s} .
\end{equation}


\subsubsection{Markovian case}
\label{markovian_case}

We consider the simplest case where the transition dynamics is
Markovian. (Even in the Markovian case, the
two-state model can reproduce some interesting dynamic properties which
reflect the dynamic heterogeneity.)
In this case, we can describe the transition dynamics between the fast and slow states
by the following master equation:
\begin{equation}
 \label{master_equation_two_state_model}
 \frac{d}{dt}
  \begin{bmatrix}
   P_{f}(t) \\
   P_{s}(t)
  \end{bmatrix}
  =
  \begin{bmatrix}
   - k_{f} & k_{s} \\
   k_{f} & - k_{s}
  \end{bmatrix}
  \cdot
  \begin{bmatrix}
   P_{f}(t) \\
   P_{s}(t)
  \end{bmatrix} .
\end{equation}
where $k_{f}$ and $k_{s}$ are the transition rates from the fast to slow
states and
from the slow to fast states, respectively.
The set of equations
\eqref{langevin_equation_two_state_model}-\eqref{master_equation_two_state_model}
can be solved analytically.

The equilibrium probabilities (equilibrium fractions) of the fast
and slow states, become
\begin{equation}
  \label{equilibrium_fraction_two_state_model}
 \phi_{f} = \frac{k_{s}}{k_{f} + k_{s}}, \qquad
 \phi_{s} = \frac{k_{f}}{k_{f} + k_{s}} .
\end{equation}
The joint probability to find the particle at the state $h'$ at time $0$ and at
the state $h$ at time $t$ (the transition probability), $W_{hh'}(t)$, can be calculated straightforwardly
from the coefficient matrix in
Eq.~\eqref{master_equation_two_state_model}. The explicit expression becomes
\begin{equation}
 \label{transition_probability_two_state_model_markovian}
   \begin{bmatrix}
    W_{ff}(t) & W_{fs}(t) \\
    W_{sf}(t) & W_{ss}(t)
   \end{bmatrix}
   =
   \begin{bmatrix}
    \phi_{f} + \phi_{s} e^{- t / \tau} &
    \phi_{f} (1 - e^{- t / \tau}) \\
    \phi_{s} (1 - e^{- t / \tau}) &
    \phi_{s} + \phi_{f} e^{- t / \tau}
   \end{bmatrix} ,
\end{equation}
where we have defined the characteristic relaxation time as $\tau \equiv 1
/ (k_{f} + k_{s})$.
The four-body two-time correlation function $\psi_{1}(t)$ can be expressed by using $W_{h
h'}(t)$ and $\phi_{h}$ as
\begin{equation}
 \label{psi1_two_state_model_markovian}
 \psi_{1}(t) = \frac{1}{D_{\text{eff}}^{2}} \sum_{h, h' = f, s}
  D_{h} D_{h'} W_{h h'}(t) \phi_{h'} - 1 .
\end{equation}
From Eqs.~\eqref{transition_probability_two_state_model_markovian} and
\eqref{psi1_two_state_model_markovian},
the explicit form of $\psi_{1}(t)$ becomes
\begin{equation}
 \label{psi1_explicit}
  \psi_{1}(t)
   = \frac{\phi_{s} \phi_{f}  (D_{f} - D_{s})^{2}}{D_{\text{eff}}^{2}}
  e^{-t / \tau} .
\end{equation}

By substituting Eq.~\eqref{psi1_explicit} into Eq.~\eqref{rsd_square_final},
finally we have a simple expression for the squard RSD:
\begin{equation}
 \label{rsd_explicit}
 \Sigma^{2}(t;\Delta)
  = \frac{\phi_{s} \phi_{f}  (D_{f} - D_{s})^{2}}{D_{\text{eff}}^{2}}
  \frac{2 \tau^{2}}{t^{2}}
  \left( e^{-t / \tau} - 1 + \frac{t}{\tau} \right) .
\end{equation}
The asymptotic forms for $t \ll \tau$ and $t \gg \tau$ are
\begin{equation}
 \label{rsd_asymptotic}
 \Sigma(t;\Delta)
  =
\begin{cases}
 \displaystyle \frac{\sqrt{\phi_{s} \phi_{f}} (D_{f} -
 D_{s})}{D_{\text{eff}}} & (t \ll \tau) , \\
 \displaystyle \frac{\sqrt{\phi_{s} \phi_{f}} (D_{f} -
 D_{s})}{D_{\text{eff}}} \sqrt{\frac{2 \tau}{t}} & (t \gg \tau) .
\end{cases}
\end{equation}
From these asymptotic forms, the crossover time $\tau_{c}$ is estimated as
$\tau_{c} = 2 \tau$. As expected, the crossover time is twice of the
relaxation time $\tau$.

\begin{figure}[h]
 \includegraphics[width=1.0\linewidth,clip]{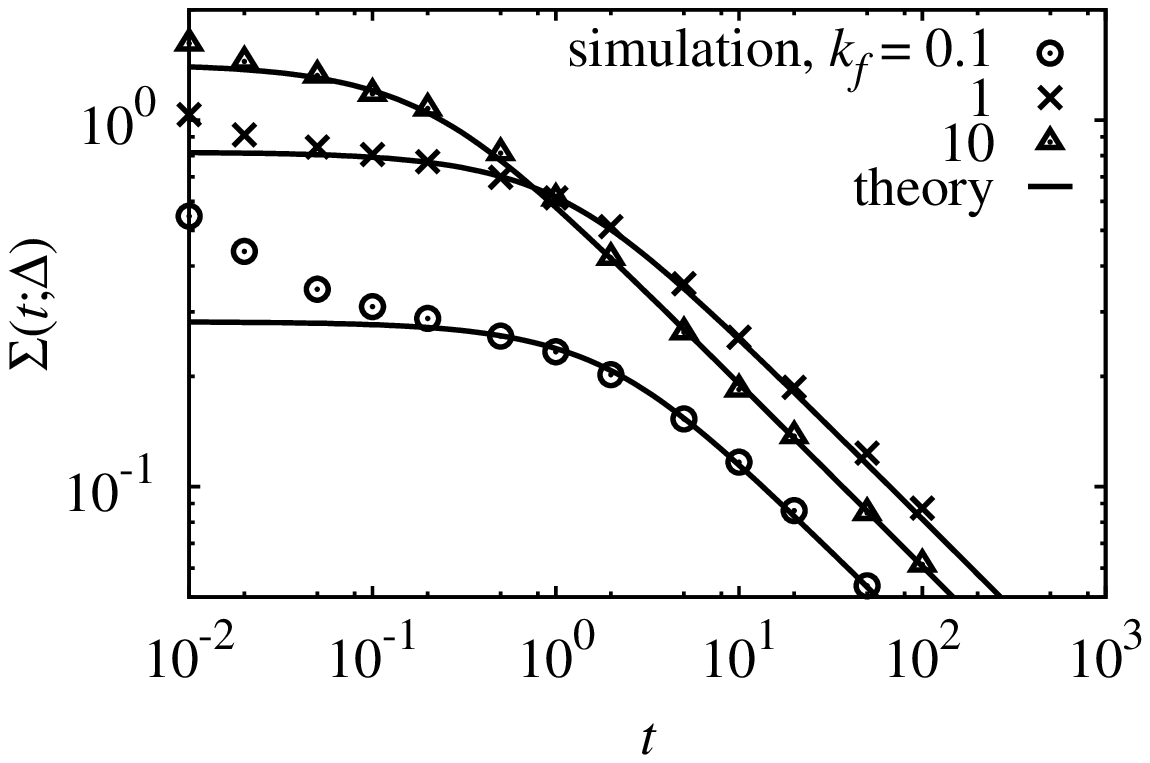}
\caption{\label{rsd_tamsd_two_state_model}}
\end{figure}

We perform the simulations and compare the simulation results with
the theoretical prediction (Eq.~\eqref{rsd_explicit} or Eq.~\eqref{rsd_asymptotic}).
We show the simulation method later, because the simulation
for the Markovian two-state model can be performed as a
special case of the non-Markovian two-state model.
We perform simulations with $D_{s} = 1$, $D_{f} = 10$, $k_{f} = 1$, and
several different values of $k_{s}$ ($k_{s} = 0.1, 1,$ and $10$).
Figure \ref{rsd_tamsd_two_state_model} shows the simulation results
together with the theoretical squared RSD (Eq.~\eqref{rsd_explicit}) and
its asymptotic forms (Eq.~\eqref{rsd_asymptotic}).
We observe that the theoretical prediction agrees well with the
simulation data, except for the small $t$ region. (The deviations at
the small $t$ region are similar to the case of the reptation model.)
Therefore, we find that our general formula (Eqs.~\eqref{rsd_square_final} and
\eqref{rsd_square_asymptotic}) can be applied to the Markovian two-state
model, where the dynamics of the instantaneous diffusivity is described
by the Markovian transition dynamics between two states.
The deviation from the
theory at the small $t$ regions is due to the correction term.
From Eq.~\eqref{rsd_square_approximated_small_t}, the
relative contribution of the correction term increases as the plateau
value of the RSD ($\psi_{1}(0)$) decreases. Actually, we observe that
the deviation is especially large for the case
of $k_{f} = 0.1$, in which the plateau value is small.

\subsubsection {Non-Markovian case}
\label{non_markovian_two_state_model}

Markovian models are varid for ideal systems where the memory effects are negligible.
If the memory effects are not negligible, the dynamics should be non-Markovian.
In this subsection, we consider the two-state model
defined by Eq.~\eqref{langevin_equation_two_state_model} with non-Markovian
transition processes between fast and slow states.
Such non-Markovian dynamics will be important when comparing the model with experimental data.
To handle non-Markovian processes, we use the renewal theoretic approach
\cite{Cox,Godreche-Luck-2001}. We assume 
that the system is initially in the equilibrium state. In other words,
we assume that the mean trapping-time does  
not diverge and the the system is well-equilibrated.
In what follows, we mainly use the same notations as the Markovian case.

We express the trapping-time distribution of the state $h$ as
$\rho_{h}(\tau)$.
Also, we express the equilibrium trapping-time distribution as 
$\rho_{h}^{\mathrm{(eq)}}(\tau)$. For example, if a particle is in
the fast state at time $t=0$, this particle became the fast state
at some time $t=t_0 <0$. If $t_1$ is the time when first transition (to
the slow state) occurs, $\tau_{1} = t_1-t_0$ obeys the distribution $\rho_f(\tau_{1})$, but $t_1$
itself does not necessarily obey $\rho_f(t_1)$.
Instead, $t_{1}$ obeys $\rho_{f}^{\mathrm{(eq)}}(\tau)$.
(We note that
the time $t_1$ is called the forward recurrence time in renewal theory
\cite{Cox}.)
The explicit expression for $\rho_{h}^{\text{(eq)}}$ is
\cite{Cox,Godreche-Luck-2001}
\begin{equation}
 \label{trapping_time_distribution_equilibrium_explicit}
  \rho^{\text{(eq)}}_{h}(\tau) = \frac{1}{\langle \tau \rangle_{h}}
  \int_{\tau}^{\infty} d\tau' \, \rho_{h}(\tau') ,
\end{equation}
where $\langle \tau \rangle_{h}$ is the average
trapping time of the state $h$, defined as
\begin{equation}
 \langle \tau \rangle_{h} \equiv \int_{0}^{\infty} d\tau \, \tau
  \rho_{h}(\tau) .
\end{equation}
(For the exponential distribution, two distributions $\rho_h(\tau)$ and $\rho_h^{\mathrm{(eq)}}(\tau)$ coincide.)
Equilibrium fractions of each state, $\phi_f$ and $\phi_s$, are given by
\begin{equation}
  \phi_f = \frac {\langle \tau \rangle_{f}}{\langle \tau \rangle_{f} + \langle\tau\rangle_{s}}, \qquad
  \phi_s = \frac {\langle \tau \rangle_{s}}{\langle \tau \rangle_{f} + \langle\tau\rangle_{s}}.
\end{equation}
For the case of the exponential trapping-time distribution,
 $\langle \tau \rangle_{h} = 1 / k_h$
and we recover Eq.~\eqref{equilibrium_fraction_two_state_model}.

The four-body two-time correlation function $\psi_{1}(t)$ can be calculated in
a similar way to the Markovian case.
Using the joint probability of being at state $h$ at time $t$, starting from at state $h'$ at time
$0$, $W_{hh'}(t)$, $\psi_{1}(t)$ can be expressed as
\begin{equation}
 \label{psi1_two_state_nonmarkov}
 \psi_{1}(t) 
  = \frac{1}{D_{\text{eff}}^{2}} \sum_{h, h' = f,s}
  D_{h} D_{h'} [W_{h h'}(t) - \phi_{h}] \phi_{h'} .
\end{equation}
Here $D_{\text{eff}}$ is the effective diffusion coefficient defined by
Eq.~\eqref{effective_diffusion_coefficient_two_state_model}.

Although it is difficult to obtain the explicit expression of
$\psi_{1}(t)$, we can obtain the asymptotic forms.
For $t = 0$, the transition probability simply becomes $W_{hh'}(0) = \delta_{hh'}$ and thus we have
\begin{equation}
 \label{psi1_limit_nonmarkov}
 \psi_1(0)
  = \frac {\phi_f\phi_s (D_f-D_s)^2}{D_{\mathrm{eff}}^2} .
\end{equation}
Eq.~\eqref{psi1_limit_nonmarkov} has formally the same form as the Markovian case
(Eq.~\eqref{psi1_explicit} with $t = 0$). This result is physically
natural because we have no transition at $t = 0$ and the details of the
transition dynamics do not affect $\psi_{1}(0)$, as long as the system
is in equilibrium.

The integral of $\psi_{1}(t)$ becomes as follows:
\begin{equation}
 \label{psi1_integral_nonmarkov}
 \int_{0}^{\infty} dt \, \psi_{1}(t) = \frac{ \phi_{f} \phi_{s} (D_{f} - D_{s})^{2}}{D_{\text{eff}}^{2}} \tilde{\tau}.
\end{equation}
Here $\tilde{\tau}$ is the characteristic relaxation time of
the non-Markovian two-state model and is defined as
\begin{equation}
 \label{relaxation_time_nonmarkov}
  \tilde{\tau} \equiv
 \left(\frac {\langle \tau^2 \rangle_{s} - \langle \tau \rangle_{s}^{2}}{\langle \tau \rangle_{s}^2}
  + \frac {\langle \tau^2 \rangle_{f} - \langle \tau \rangle_{f}^{2}}{\langle \tau \rangle_{f}^2} \right)
  \frac {\langle \tau \rangle_{f} \langle \tau \rangle_{s}}
  {2(\langle \tau \rangle_{f} + \langle \tau \rangle_{s})} .
\end{equation}
(For the exponential trapping-time distribution, we simply have
$\tilde{\tau} = \tau$.) The detailed calculations for Eq.~\eqref{psi1_integral_nonmarkov}
are summarized in Appendix
\ref{detailed_calculations_for_non_markovian_two_state_model}.

From Eqs.~\eqref{rsd_square_asymptotic},
\eqref{psi1_limit_nonmarkov}, and
\eqref{psi1_integral_nonmarkov}, we have the following asymptotic forms
for the RSD:
\begin{equation}
 \label{rsd_square_asymptotic_non_markovian}
  \Sigma(t;\Delta)
  \approx
  \begin{cases}
   \displaystyle
   \frac {\sqrt{\phi_f\phi_s} (D_f-D_s)}{D_{\mathrm{eff}}} & (t \ll \tilde{\tau}), \\
   \displaystyle
   \frac {\sqrt{\phi_f\phi_s} (D_f - D_s)}{D_{\mathrm{eff}}}
  \sqrt{\frac {2\tilde{\tau}}{t}}
    & (t \gg \tilde{\tau}) .
  \end{cases}
\end{equation}
Eq.~\eqref{rsd_square_asymptotic_non_markovian} has almost the same form
as the Markovian case,
Eq.~\eqref{rsd_asymptotic} ($\tau$ in Eq.~\eqref{rsd_asymptotic} is replaced by $\tilde{\tau}$).
Thus our theory predicts similar crossover behavior as the Markovian
case. The crossover time is estimated as $\tau_c = 2
\tilde{\tau}$. Here it should be emphasized that the correlation function $\psi_{1}(t)$ of the non-Markovian two-state model
is not a single exponential form.
The crossover time depends on the average relaxation time $\tilde{\tau}$
defined in Eq.~\eqref{relaxation_time_nonmarkov}, and thus is not a
simple arithmetic nor harmonic averages of $\langle \tau \rangle_{f}$ and
$\langle \tau \rangle_{s}$.
Eq.~\eqref{rsd_square_asymptotic_non_markovian} gives only the
asymptotic forms for the RSD. The detailed transition behavior from the constant RSD
($\Sigma(t;\Delta) \propto t^{0}$) to the Gaussian decay
($\Sigma(t;\Delta) \propto t^{-1/2}$) can be qualitatively different from the Markovian case.

To examine the validity of Eq.~\eqref{rsd_square_asymptotic_non_markovian}, we perform
simulations for the non-Markovian two-state model and compare the
simulation results with Eq.~\eqref{rsd_square_asymptotic_non_markovian}.
The simulation scheme consists of two steps. First, we
sample the waiting time $\tau$ at the current state from the waiting
time distribution. Second, we integrate the Langevin equation
\eqref{langevin_equation_two_state_model} until the sampled waiting
time. Then we change the state and go back to the first step and iterate
the same procedure. (This simulation scheme can be applied for the
Markovian two-state model, if we use the exponential distribution
functions both for the fast and slow states.)

\begin{figure}[h]
 \includegraphics[width=1.0\linewidth,clip]{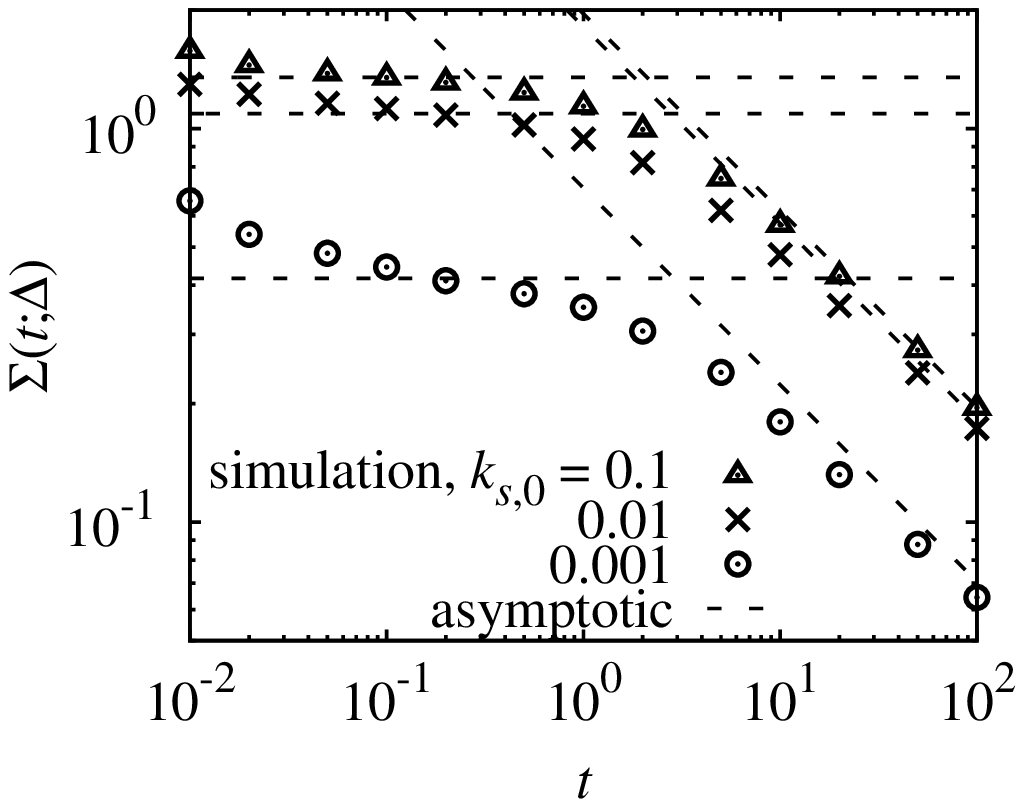}
\caption{\label{rsd_tamsd_two_state_model_nonmarkov}}
\end{figure}

In this work, we employ the exponential distribution for the fast state
and a non-exponential distribution for the slow state, as follows:
\begin{align}
 & \label{model_waiting_time_fast_non_markovian}
 \rho_{f}(\tau) = k_{f} e^{-k_{f} \tau}, \\
 & \label{model_waiting_time_slow_non_markovian}
 \rho_{s}(\tau) = \int_{k_{s,0}}^{k_{s,1}} dk_{s} \, k_{s} e^{-k_{s} \tau} q_{s}(k_{s}).
\end{align}
Here $k_{s,0}$ and $k_{s,1}$ are the lower and upper limits for the
transition rate at the slow state, and $q_{s}(k_{s})$ is the distribution of
the transition rate.
We employ the following power-law type distribution for $q_{s}(k_{s})$,
\begin{equation}
 q_{s}(k_{s}) = \frac{\alpha - 1}{k_{s,1}^{\alpha - 1} - k_{s,0}^{\alpha
  - 1}} {k_s}^{\alpha - 2} ,
\end{equation}
with $0 \le \alpha \le 1$ being the power-law exponent. As a result, the distribution for the slow state obeys a power law for small 
$\tau$ and exponential distribution for large $\tau$. (The transition
from a power law to the exponential occurs at $\tau \approx k_{s,0}^{-1}$.)
We set parameters as $D_{s} = 1$, $D_{f} = 10$, $k_{f} = 1$, $k_{s,1} =
1$, $\alpha = 0.2$. We vary the value of $k_{s,0}$ as $k_{s,0} = 0.1,
0.01,$ and $0.001$, to control the non-Markovian transition dynamics.
Figure \ref{rsd_tamsd_two_state_model_nonmarkov} shows the simulation
results together with the theoretically derived asymptotic forms.
The simulation data show clear crossovers for the RSD, as the case of the Markovian
two-state model.
We observe the asymptotic forms by our theory (Eq.~\eqref{rsd_square_asymptotic_non_markovian}) agree with
the simulation data. 

However, because $\psi_1(t)$ is not a single exponential form and has
rather broad distribution of relaxation times, the crossover region
becomes broad compared with the Markovian case.
This means that there is some deviations from the
asymptotic form Eq.~\eqref{rsd_square_asymptotic}, at the intermediate
$t$ region.
Especially for the case of $k_{s,0} = 0.001$, the deviation from two
asymptotic forms is relatively large. (Also, as the case of the
Markovian case, the contribution of the
correction term is relatively large for $k_{s,0} = 0.001$. This is
another reason why the deviation from the theoretical form is relatively
large for $k_{s,0} = 0.001$.)

\section {Discussions}
\subsection {Comparison with Other Analysis Methods}

We have shown that our general formula for the RSD of the TAMSD works
well for several analytically solvable systems.
Here we compare our
analysis method with other methods. For supercooled liquids, so far,
several different quantities have been employed to analyze the dynamic heterogeneity.

Yamamoto and Onuki
\cite{Yamamoto-Onuki-1998,Yamamoto-Onuki-1998a} showed that the van Hove correlation
function can resolve the dynamic heterogeneity. The van Hove self-correlation
function is defined as
\begin{equation}
 G_{s}(\bm{r},\Delta) \equiv \langle \delta(\bm{r} - \bm{r}(\Delta) +
  \bm{r}(0)) \rangle .
\end{equation}
For a relatively short time scale, $G_{s}(\bm{r},\Delta)$ shows non-Gaussian
behavior, due to the dynamic heterogeneity. For a
relatively long time scale, $G_{s}(\bm{r},\Delta)$ approaches to the Gaussian
behavior, which corresponds to the ergodic state. Although the van Hove
correlation is useful to qualitatively observe the dynamic
heterogeneity, it is not easy to quantitatively determine, for example, the
crossover time directly from the van Hove correlation function. For such
a purpose, scalar quantities are preferred than distribution functions.
To quantify the non-Gaussian behavior, so called the non-Gaussianity parameter has been
utilized. The non-Gaussianity parameter is defined as \cite{Rahman-1964,Ernst-Kohler-Weiss-2014,Cherstvy-Metzler-2014}
\begin{equation}
 \label{non_gaussianity_parameter_definition}
  A(\Delta) \equiv \frac{n \langle [\bm{r}(\Delta) -
  \bm{r}(0) ]^{4} \rangle}{(n + 2) \langle [\bm{r}(\Delta) -
  \bm{r}(0)]^{2}\rangle^{2}} - 1 .
\end{equation}
This parameter becomes non-zero if the distribution of the displacement
(the van Hove correlation function) is not Gaussian.
Although the
non-Gaussianity parameter can characterize the long time relaxation behavior,
its explicit expression for the time-dependent diffusivity model is not
simple compared with our general formula for the RSD, as shown in
Appendix \ref{non_gaussianity_parameter}.

Recent simulation and theoretical works show that the four-point
time-space correlation function is an important quantity in supercooled
liquids \cite{Glotzer-Novikov-Schroder-2000,Lacevic-Starr-Schroder-Glotzer-2003,Berthier-Biroli-2011,Mizuno-Yamamoto-2011}.
The four-point dynamic correlation function is defined as
\begin{equation}
 \begin{split}
  \chi_{4}(\bm{r},\bm{r}',\Delta)
  \equiv &
  \langle \delta\rho(\bm{r},\Delta) \delta\rho(\bm{r},0)
  \delta\rho(\bm{r}',\Delta) \delta\rho(\bm{r}',0) \rangle \\
  & - \langle \delta\rho(\bm{r},\Delta) \delta\rho(\bm{r},0) \rangle
 \langle \delta\rho(\bm{r}',\Delta) \delta\rho(\bm{r}',0) \rangle ,
 \end{split}
\end{equation}
where $\delta\rho(\bm{r},t)$ is the density fluctuation at position $\bm{r}$
and time $t$. The four-point correlation function can also quantify the dynamic
heterogeneity and was analyzed in detail in recent works.
Although the RSD of the TAMSD is not equivalent to the four-point correlation
function, nor the non-Gaussianity parameter, the RSD of the TAMSD can be
utilized in a similar way to these quantities.
As far as the
authors know, the RSD or RF analysis is not performed for MD data of
supercooled liquids. 
The application of the TAMSD analysis to the MD
simulation data of the supercooled liquids is an interesting future
work.
In particular, the comparison of the crossover time determined from
the RSD of the TAMSD with other characteristic times (such as the
$\alpha$-relaxation time) will be interesting.

Garrahan, Chandler and coworkers
\cite{Hedges-Jack-Garrahan-Chandler-2009,Chandler-Garrahan-2010,Biroli-Garrahan-2013}
analyzed so-called the ``activity'' to study
the dynamics of supercooled liquids. The activity is defined as:
\begin{equation}
 \label{activity_definition}
 K[x] \equiv \Delta \sum_{t' = 0}^{t}
  [\bm{r}(t' + \Delta) - \bm{r}(t')]^{2} .
\end{equation}
Here $x(t)$ represents the point in the configuration space, and $\Delta
$ and $t$ are the time step size and the observation time,
respectively. $\bm{r}(t)$ represents the trajectory of a particle and it
depends on $x(t)$.
The summation over $t'$ in Eq.~\eqref{activity_definition} represents the sum
taken for every
$\Delta$, within the observation time window ($0 < t' < t$).
The activity is essentially the same as the TAMSD.
(The summation over $t'$
may be replaced by the integral over $t'$, and then the activity
reduced to the TAMSD except the normalization factor $1 / (t -
\Delta)$.)
Thus the activity can be interpreted as a
stochastic variable just like the TAMSD. 
(The activity is a time-averaged quantity but explicitly depends on the position in the
configuration space. Naively, we expect that $x(t)$ contains the same
information as $\bm{B}(t)$ in our model.)
Hedges et al
\cite{Hedges-Jack-Garrahan-Chandler-2009} showed that the ensemble
average of the activity can be utilized as the order parameter and the
glass transition can be interpreted in analogy to the first order phase
transition. Although their approach is different from ours, we consider
that the fluctuation analysis of the TAMSD in this work can also provide useful
information for the dynamics of supercooled liquids.

There are other analysis methods which do not utilize the MSD.
For example, to analyze the longest relaxation time in entangled
polymers, the relaxations of the stress and the end-to-end vector
are simple and useful\cite{Doi-Edwards-book}.
As already pointed in the previous
work\cite{Uneyama-Akimoto-Miyaguchi-2012}, the RF and RSD analyses give
qualitatively similar long time relaxation behavior as other analysis
methods. Our general formula (Eq.~\eqref{tau_c_explicit}) or the 
analytic result for the reptation model (Eq.~\eqref{tau_c_explicit_reptation})
gives the relation between the relaxation time distribution and the
crossover time $\tau_{c}$. In general, if the relaxation is not a single
exponential type, $\tau_{c}$ becomes quantitatively
different from the characteristic relaxation times determined by other
analysis methods. Thus, the comparison of $\tau_{c}$ with other
relaxation time data can provide the information on the relaxation time
distribution. For example, in the case of entangled polymers, the ratio
$\tau_{c} / \tau_{d}$ ($\tau_{d}$ is determined from the stress
relaxation) can be utilized as an index for the contribution of
non-reptation type relaxation mechanisms.

{
The advantage of the RSD analysis is that it directly reflects the
dynamics of the instantaneous diffusivity. From
Eq.~\eqref{rsd_square_final}, the RSD can be directly related to the
correlation function $\psi_{1}(t)$. From Eqs.~\eqref{correlation_fucntion_psi1_definition} and
\eqref{rsd_square_final}, we have the following relation:
\begin{equation}
 \label{time_correlation_diffusivity_from_squared_rsd}
 \frac{\langle \tr \bm{D}(t) \tr \bm{D}(0) \rangle}{\langle \tr \bm{D}
  \rangle^{2}}
  \approx \frac{1}{2} \frac{\partial^{2}}{\partial t^{2}} \left[ t^{2}
  \Sigma^{2}(t;\Delta) \right] .
\end{equation}
Eq.~\eqref{time_correlation_diffusivity_from_squared_rsd} means that if
we have the RSD of the TAMSD for several different observation times, we
can calculate the correlation function for the time-depending and
fluctuating diffusivity. As far as the authors know, there is
no such analysis method which gives the correlation function of the
diffusivity. Eq.~\eqref{time_correlation_diffusivity_from_squared_rsd} will be
especially useful for the
analysis of experimental data, because we cannot directly observe the
diffisuvity from the trajectories.
}

\subsection {Time-Dependent Diffusivity Model and Other Models}

As shown in Section~\ref{model}, various dynamics models can be expressed as the
Langevin equation with time-dependent and fluctuating diffusivity. Here we discuss
the relation between the time-dependent diffusivity model (described in
Section~\ref{model}) and other
dynamics models.

{\L}uczka, Niemiec and Piotrowski
\cite{Luczka-Niemiec-Piotrowski-1992,Luczka-Niemiec-Piotrowski-1993}
considered the randomly interrupted diffusion model, in which the
strength of the noise in the Langevin equation depends on another
stochastic process. The time-dependent diffusivity model reduces to the
randomly interrupted diffusion model by tuning the dynamics of the noise
coefficient matrix.
Fogedby \cite{Fogedby-1994} considered two coupled Langevin equations. Fogedby replaced the
time in a usual Langevin equation by the virtual time, and introduced
another Langevin equation for the evolution of the virtual time. The virtual
time may be interpreted as the time-dependent and fluctuating diffusivity.
Thus, we can interpret the Fogedby model as a special case of the
time-dependent and fluctuating diffusivity model.
However, we should note that the Fogedby model is designed to reproduce
the L\'{e}vy flight, and thus the dynamics of the 
virtual time is assumed to be non-ergodic, which is different from our model.
Recently, Jeon, Chechkin and Metzler \cite{Jeon-Chechkin-Metzler-2014}
considered a time-dependent diffusion coefficient model. In their model,
the diffusion coefficient simply depends on time $t$ as $D(t) \propto t^{\alpha
- 1}$ (with $\alpha$ being an exponent). Namely, the dynamics of the instantaneous diffusion coefficient
matrix is deterministic. Such a dynamics model reproduces
the anomalous diffusion behavior. Using non-ergodic dynamics to the noise
coefficient matrix or the instantaneous diffusion coefficient matrix, we have 
anomalous diffusion in the time-dependent and fluctuating diffusivity model.

When the noise coefficient matrix or the instantaneous diffusion coefficient matrix
obeys the discrete jump dynamics, the diffusion behavior strongly
depends to the properties of the jump dynamics (as shown for the
non-Markovian two-state model in Section \ref{non_markovian_two_state_model}). Such jump dynamics is
often modeled as the continuous-time random walk (CTRW) \cite{Metzler-Klafter-2000}. The CTRW is
used, for example, as the diffusion model on the random potential
landscape. Klafter and Silbey derived the CTRW for the diffusion model on randomly occupied
lattices, by using the projection operator technique
\cite{Klafter-Silbey-1980}. Here we show that the two-state model
reduces to the CTRW at a certain limit.

We start from 
the non-Markovian two-state model. In general, the non-Markovian
two-state model does not reduce to the CTRW, although some aspects of
the model are similar to the CTRW. We consider the special case where
$D_{s} = 0$.
In this case, the particle does not move when it is in
the slow state. The particle can move freely in the fast
state whereas the particle is trapped and cannot move in the slow state.
If the average sojourn time in the fast state, $\langle \tau \rangle_{f}$, is very short, the
movement of the particle looks like the instantaneous and discrete jump. Thus, at
the limit of $\langle \tau_{f} \rangle \to 0$ with $D_{f} \langle
\tau_{f} \rangle = \text{(const.)}$, the trajectory of the Brownian
particle reduces to that of the CTRW. The step size distribution for the CTRW is determined from
the sojourn time distribution and the diffusion coefficient of the fast state.
The trapping-time
distribution for the slow state directly corresponds to the waiting time
distribution for the CTRW.
This can be interpreted as a simple and complementary derivation of the CTRW from
a microscopic dynamics model.

It would be informative to mention a connection between the Langevin
equation with the time-dependent and fluctuating diffusivity and the
CTRW, from the view point of the fluctuation analysis.
If we take the limit of $\langle \tau_{f} \rangle \to 0$,
the crossover of the RSD disappears because the crossover time $\tau_{c} =
2 \tilde{\tau}$
goes to zero at this limit.
Actually, in the CTRW, the RSD does not show the plateau at the short time region.
(The constant RSD can be observed only for nonequilibrium initial ensembles.)
However, the RF and RSD in the CTRW \cite{Miyaguchi2011a,Miyaguchi-Akimoto-2013}
also show the crossover behavior somewhat similar to one in the Langevin equation in
this work.
As we mentioned, the crossover time $\tau_{c}$ goes to zero at the
limit, and thus this crossover behavior of the RSD in the CTRW has a qualitatively different
origin from one in the Langevin equation with the time-dependent and
fluctuating diffusivity. The crossover in the CTRW is related to the
cutoff time of the waiting time distribution\cite{Miyaguchi2011a,Miyaguchi-Akimoto-2013}, not to $\tau_{c}$ in our analysis.
When the waiting time distribution in the CTRW obeys a power law with an exponential cutoff,
the RSD in the short $t$ region shows a power-law type behavior $\Sigma(t;\Delta) \propto
t^{-(1-\alpha)/2}$ (with $\alpha > 0$ being the power-law
exponent). This behavior reflects the information on the trapping-time
distribution, and such a behavior is not considered in the analysis in this work.
A power-law type behavior of the RSD will be observed in our Langevin model, 
if the waiting time has the power-law form in a rather wide range
($k_{s,1} / k_{s,0} \gg 1$ in Eq.~\eqref{model_waiting_time_slow_non_markovian}).
In fact, as shown in Figure~\ref{rsd_tamsd_two_state_model_nonmarkov}, the crossover behavior becomes rather broad
in the case of $k_{s,0} = 0.001$. Such $t$-dependence is somewhat
similar to one observed in the CTRW.

\section {Conclusions}
\label{conclusions}

We have derived the formula for the RSD of the TAMSD (which quantifies the
fluctuation of the TAMSD) as a function of the observation time, 
in the Langevin equation with the time-dependent and fluctuating
diffusivity.
From the asymptotic behavior, a crossover from a constant RSD ($\Sigma(t;\Delta) \propto
t^{0}$) to a Gaussian decay ($\Sigma(t;\Delta) \propto t^{-1/2}$), is
predicted if there is a characteristic relaxation time of the fluctuating
diffusivity. The asymptotic forms of our formula give
the relation between the crossover time and the relaxation time.
The crossover time is given as the weighted average
relaxation time for the fluctuating diffusivity. Such a characteristic time cannot be calculated
from the EAMSD.
Applying the formula to the reptation 
model and the two-state models, we have shown that the crossover time
can actually characterize the relaxation times of the diffusivities.
This is becasue RSD reflects the dynamcis of the time-dependent and fluctuating
diffusivity. Our result
justifies our previous study \cite{Uneyama-Akimoto-Miyaguchi-2012} in
which we have numerically found that the crossover time can characterize
the long time relaxation behavior.
We also showed that the (non-Markovian) two-state model reduces to the
CTRW at a certain limit. However, this does not mean that the Langevin equation with the time-depending and
fluctuating diffusivity is equivalent to the CTRW. Actually, the
behavior of the RSD of the CTRW is qualitatively different from one of
the Langevin equation model.
The RSD analysis extracts important information for underlying
fluctuating diffusion processes. The RSD can be directly related to the
correlation function of the instantaneous diffusivity,
which is difficult to  directly extract from the trajectories.
We expect the analysis of the RSD of the TAMSD is also useful for more
complex systems such as MD simulations for
entangled polymers and supercooled liquids, single-particle-tracking
experiments, and diffusion in confined systems \cite{Akimoto2015}.
The RSD analysis together with other analysis methods will also give important
information.

\section* {Acknowledgment}

TU thanks Prof. Koh-hei Nitta for helpful comments and supports.
The authors thank Eli Barkai, Juan P. Garrahan, and Ralf Metzler for helpful comments
and informing some information on their publications.
TU was supported by Grant-in-Aid (KAKENHI) for Young
Scientists B 25800235.
TA was supported by Grant-in-Aid (KAKENHI) for Young
Scientists B 26800204.

\appendix

\section{Detailed Calculations for Ensemble Average of Squared Time-Averaged Mean Square Displacement}
\label{detailed_calculations_for_ensemble_average_of_squared_time-averaged_mean_square_displacement}

In this appendix, we show the detailed calculations for
the ensemble average of the squared TAMSD, $\langle [\overline{
\delta^{2}(\Delta;t)} ]^{2} \rangle$. From Eq.~\eqref{tamsd_definition}, $\langle [\overline{
\delta^{2}(\Delta;t)} ]^{2} \rangle$ can be explicitly written in terms
of the noise $\bm{w}(t)$ and the noise coefficient matrix $\bm{B}(t)$:
\begin{equation}
 \label{tamsd_square_average_initial}
  \begin{split}
   \langle [\overline{ \delta^{2}(\Delta;t)} ]^{2} \rangle
   & = \frac{2}{(t - \Delta)^{2}} \int_{0}^{t - \Delta} dt' \int_{0}^{t'} dt''
   \, \langle [\bm{r}(t' + \Delta) - \bm{r}(t')]^{2}
    [\bm{r}(t'' + \Delta) - \bm{r}(t'')]^{2} \rangle \\
   & = \frac{8}{(t - \Delta)^{2}} \int_{0}^{t - \Delta} dt' 
   \int_{0}^{t'} dt'' \int_{t'}^{t' + \Delta} ds \int_{t'}^{t' + \Delta} ds'
     \int_{t''}^{t'' + \Delta} du \int_{t''}^{t'' + \Delta} du' \\
   & \qquad \times
   \langle w_{i}(s) w_{j}(s') w_{k}(u) w_{l}(u') \rangle
   \langle B_{mi}(s) B_{mj}(s') B_{nk}(u) B_{nl}(u') \rangle .
  \end{split}
\end{equation}
Here we have employed the Einstein summation convention.
By utilizing the Wick's theorem\cite{Doi-Edwards-book}, Eq.~\eqref{tamsd_square_average_initial} can be rewritten as
\begin{equation}
 \label{tamsd_square_average}
  \begin{split}
   \langle [\overline{ \delta^{2}(\Delta;t)} ]^{2} \rangle
   & = \frac{8}{(t - \Delta)^{2}} \int_{0}^{t - \Delta} dt' 
   \int_{0}^{t'} dt'' \bigg[ \int_{t'}^{t' + \Delta} ds 
     \int_{t''}^{t'' + \Delta} du \, \langle \tr \bm{D}(s) \tr \bm{D}(u) \rangle \\
   & \qquad + 2 \int_{t'}^{t' + \Delta} ds 
     \int_{t''}^{t'' + \Delta} du
    \int_{t'}^{t' + \Delta} ds' \int_{t''}^{t'' + \Delta} du' \,
   \delta(s - u') \delta(s' - u)
   \tr \langle \bm{D}(s) \cdot \bm{D}(u) \rangle \bigg] .
  \end{split}
\end{equation}
The integrals over $s$, $s'$, $u$ and $u'$ in the last line of
Eq.~\eqref{tamsd_square_average} can be calculated as follows. For an arbitrary
function $f(s,s',u,u')$, we have the following relation for the integrals over $s$ and $u'$:
\begin{equation}
 \int_{t'}^{t' + \Delta} ds \int_{t''}^{t'' + \Delta} du' \,
   \delta(s - u') f(s,s',u,u') =
  \begin{cases}
   \displaystyle
   \int_{t'}^{t'' + \Delta} ds \, f(s,s',u,s)
   & (t'' + \Delta \ge t') , \\
   \displaystyle 0 & (t'' + \Delta < t') .
  \end{cases}
\end{equation}
Here we have utilized the condition $t' > t''$, which holds for the
integrand in \eqref{tamsd_square_average}.
Then the integrals in Eq.~\eqref{tamsd_square_average} become
\begin{equation}
 \label{second_integral_tamsd_square_average}
  \begin{split}
   & \int_{t'}^{t' + \Delta} ds 
     \int_{t''}^{t'' + \Delta} du
    \int_{t'}^{t' + \Delta} ds' \int_{t''}^{t'' + \Delta} du' \,
   \delta(s - u') \delta(s' - u)
   \tr \langle \bm{D}(s) \cdot \bm{D}(u) \rangle \\
   & = \begin{cases}
   \displaystyle
   \int_{t'}^{t'' + \Delta} ds  \int_{t''}^{t'' + \Delta} du
    \int_{t'}^{t' + \Delta} ds' \, \delta(s' - u)
   \tr \langle \bm{D}(s) \cdot \bm{D}(u) \rangle
   & (t'' + \Delta \ge t') , \\
   \displaystyle 0 & (t'' + \Delta < t') ,
     \end{cases} \\
   & = \begin{cases}
   \displaystyle
   \int_{t'}^{t'' + \Delta} ds  \int_{t'}^{t'' + \Delta} du \,
   \tr \langle \bm{D}(s) \cdot \bm{D}(u) \rangle
   & (t'' + \Delta \ge t') , \\
   \displaystyle 0 & (t'' + \Delta < t') .
     \end{cases}
  \end{split}
\end{equation}
By using Eq.~\eqref{second_integral_tamsd_square_average}, Eq.~\eqref{tamsd_square_average} can be rewritten as
\begin{equation}
 \label{tamsd_square_average_modified}
  \begin{split}
   \langle \overline{ [\delta^{2}(\Delta;t)}]^{2} \rangle
   & = \frac{8}{(t - \Delta)^{2}} \int_{0}^{t - \Delta} dt' 
   \int_{0}^{t'} dt'' \int_{t'}^{t' + \Delta} ds 
     \int_{t''}^{t'' + \Delta} du \, \langle \tr \bm{D}(s) \tr \bm{D}(u)
   \rangle \\
   & \qquad + \frac{16}{(t - \Delta)^{2}} \int_{0}^{t - \Delta} dt' 
   \int_{0}^{t'} dt'' \, \Theta(t'' - t' + \Delta)
   \int_{t'}^{t'' + \Delta} ds 
   \int_{t'}^{t'' + \Delta} du \,
   \tr \langle \bm{D}(s) \cdot \bm{D}(u) \rangle \\
   & = \frac{8}{(t - \Delta)^{2}} \int_{0}^{t - \Delta} dt' 
   \int_{0}^{t'} dt'' \int_{t'}^{t' + \Delta} ds 
     \int_{t''}^{t'' + \Delta} du \, \langle \tr \bm{D}(s) \tr \bm{D}(u)
   \rangle \\
   & \qquad + \frac{32}{(t - \Delta)^{2}} \int_{0}^{t - \Delta} dt' 
   \int_{\max(0,t' - \Delta)}^{t'} dt'' 
   \int_{t'}^{t'' + \Delta} ds 
   \int_{t'}^{s} du \,
   \tr \langle \bm{D}(s) \cdot \bm{D}(u) \rangle ,
  \end{split}
\end{equation}
with $\Theta(t)$ being the Heaviside step function.
This gives Eq.~\eqref{tamsd_square_average_exact}.

\section {Relation between Relative Fluctuation and Relative Standard Deviation}
\label{relation_between_relative_fluctuation_and_relative_standard_deviation}

In this appendix, we consider the relation between the RF and the RSD
for the TAMSD. Due to the nature of the absolute value, the analytic
treatment of the RF is not easy compared with the RSD. We consider
two asymptotic limits of the RF, which can be calculated straightforwardly.

As the case of the calculation for the RSD, we assume $\Delta \ll t$.
For the small $t$ case ($t \ll \tau$)
the RSD becomes constant as given by Eq.~\eqref{rsd_square_asymptotic}. It can be rewritten as
\begin{equation}
 \label{rsd_asymptotic_small}
 \Sigma(t;\Delta) \approx \sqrt{\psi_{1}(0)} = \frac{\sqrt{\langle (\tr \bm{D})^{2}
  \rangle - (\tr \langle \bm{D} \rangle)^{2}}}{\tr \langle \bm{D}
  \rangle}.
\end{equation}
From Eq.~\eqref{rsd_asymptotic_small}, we find that
$\Sigma(t;\Delta)$ is expressed as the relative standard
deviation of $\tr \bm{D}$ for the equilibrium distribution.
This can be understood as follows.
In the case of $t \ll \tau$, we can approximate the instantaneous
diffusion coefficient matrix by its initial value $\bm{D}(0)$, and 
the TAMSD of each realization can be reasonably approximated as
\begin{equation}
 \label{tamsd_short_time_approximation}
 \overline{\delta^{2}(\Delta;t)} \approx 2 \tr \bm{D}(0)
  \Delta .
\end{equation}
If we use Eq.~\eqref{tamsd_short_time_approximation}, the RSD of the
TAMSD can be approximated as the RSD of $2 \tr \bm{D}(0) \Delta$, which
is equivalent to Eq.~\eqref{rsd_asymptotic_small}.
In a similar way, the RF can be approximately expressed as the RF of
$2 \tr \bm{D}(0) \Delta$. Thus we have the following expression
for the RF:
\begin{equation}
 \label{rf_asymptotic_small}
 R(t;\Delta) \approx \frac{\langle |\tr \bm{D} - \tr \langle \bm{D}
  \rangle|  \rangle}{\tr \langle \bm{D} \rangle} .
\end{equation}
Unfortunately, we cannot calculate $R(t;\Delta)$ further without the explicit form of
the equilibrium distribution for $\tr \bm{D}$. Nevertheless, we can formally relate
$R(t;\Delta)$ to $\Sigma(t;\Delta)$ as
\begin{equation}
 \label{rf_asymptotic_small_modified}
 R(t;\Delta) \approx \frac{\langle |\tr \bm{D} - \tr \langle \bm{D}
  \rangle|  \rangle}{\sqrt{\langle (\tr \bm{D})^{2}
  \rangle - (\tr \langle \bm{D} \rangle)^{2}}} \Sigma(t;\Delta) .
\end{equation}

For the large $t$ case ($t \gg \tau$), we have
the relation $\Sigma(t;\Delta) \propto t^{-1/2}$ from
Eq.~\eqref{rsd_square_asymptotic}. This means
that the distribution of the TAMSD is given as a Gaussian with the aid
of the central limit theorem.
We can explicitly write the
distribution of the TAMSD as follows:
\begin{equation}
 \label{tamsd_distribution_gaussian}
 P\left(\overline{\delta^{2}(\Delta;t)} \right)
  \approx \frac{1}{\sqrt{2 \pi} \, \Sigma (t;\Delta) \langle \overline{\delta^{2}(\Delta;t)} \rangle } \exp
  \bigg[ - \frac{[ \overline{\delta^{2}(\Delta;t)} -  \langle \overline{\delta^{2}(\Delta;t)} \rangle ]^{2}}
{2 [\Sigma(t;\Delta)  \langle \overline{\delta^{2}(\Delta;t)} \rangle]^{2}}
  \bigg] .
\end{equation}
The RF can be then calculated to be
\begin{equation}
 \label{rf_asymptotic_large}
 \begin{split}
  R(t;\Delta)
  & \approx \frac{1}{\langle \overline{\delta^{2}(\Delta;t)} \rangle}
  \int d\overline{\delta^{2}(\Delta;t)} \,
  \left| \overline{\delta^{2}(\Delta;t)} - \langle \overline{\delta^{2}(\Delta;t)} \rangle \right|
  \, P\left(\overline{\delta^{2}(\Delta;t)} \right)\\
  & \approx \sqrt{\frac{2}{\pi}} \Sigma(t;\Delta) .
 \end{split}
\end{equation}
By combining Eqs.~\eqref{rf_asymptotic_small_modified} and \eqref{rf_asymptotic_large},
we have the asymptotic forms of $R(t;\Delta)$:
\begin{equation}
 \label{rf_asymptotic}
 R(t;\Delta) \approx
  \begin{cases}
   \displaystyle \frac{\langle |\tr \bm{D} - \tr \langle \bm{D}
  \rangle|  \rangle}{\sqrt{\langle (\tr \bm{D})^{2}
  \rangle - (\tr \langle \bm{D} \rangle)^{2}}} \Sigma(t;\Delta) & (t \ll \tau) , \\
   \displaystyle \sqrt{\frac{2}{\pi}} \Sigma(t;\Delta) & (t \gg \tau) .
  \end{cases}
\end{equation}
From Eq.~\eqref{rf_asymptotic}, we find that the RF behaves qualitatively
in the same way as the RSD.
The crossover time determined by the RF,
$\tau_{c}'$, is different from $\tau_{c}$ in Eq.~\eqref{tau_c_explicit} by a constant factor:
\begin{equation}
 \tau_{c}' = \frac{2 [ \langle (\tr \bm{D})^{2}
  \rangle - (\tr \langle \bm{D} \rangle)^{2} ]}{\pi \langle |\tr \bm{D} - \tr \langle \bm{D}
  \rangle|  \rangle^{2}}
 \tau_{c} .
\end{equation}
In most practical cases, the ratio of $\tau_{c}'$ and $\tau_{c}$ is
of the order of unity, and thus both $R(t;\Delta)$ and
$\Sigma(t;\Delta)$ can be utilized to analyze the long time relaxation behavior.

\begin{figure}[p]
 \includegraphics[width=1.0\linewidth,clip]{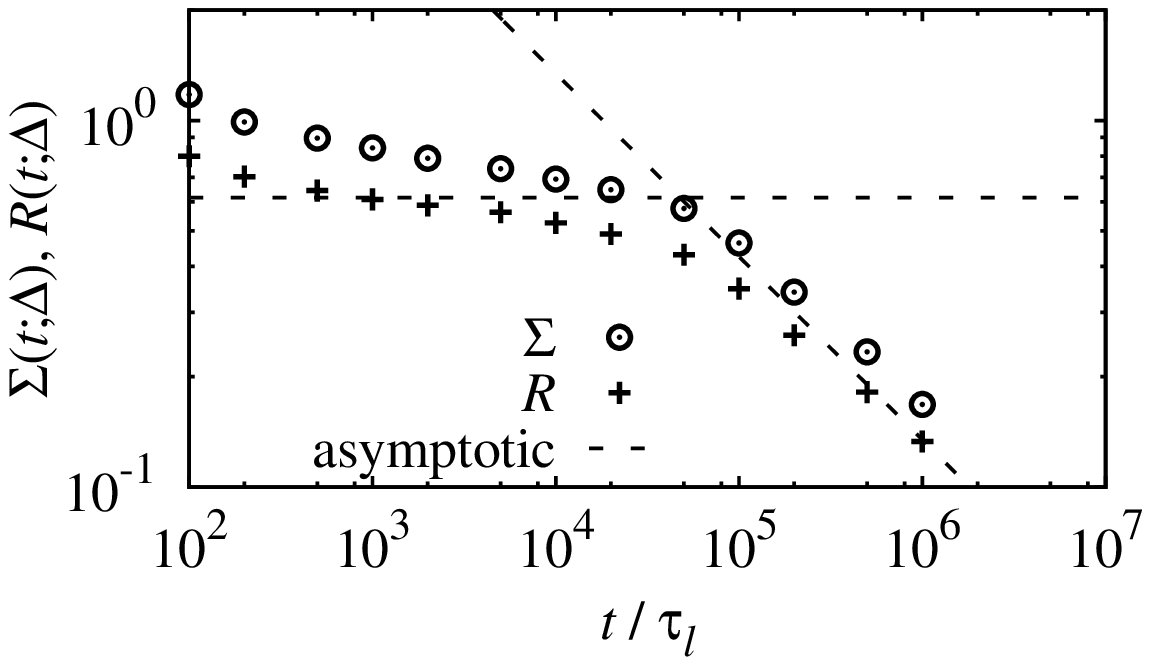}
\caption{\label{rsd_and_rf_tamsd_reptation}}
\end{figure}

In the case of the reptation model,
the explicit asymptotic forms can be calculated from the equilibrium
distribution for the end-to-end vector $\bm{p}$ \cite{Doi-Edwards-book}:
\begin{equation}
 P^{(\text{eq})}(\bm{p}) = \left(\frac{3}{2 \pi \langle \bm{p}^{2} \rangle}\right)^{3/2}
  \exp \left( - \frac{3 \bm{p}^{2}}{2 \langle \bm{p}^{2} \rangle}\right) .
\end{equation}
The asymptotic forms of $\Sigma(t;\Delta)$ is given by Eq.~\eqref{rsd_asymptotic_reptation}.
Finally we have the following asymptotic forms for $R(t;\Delta)$:
\begin{equation}
 \label{rf_asymptotic_reptation}
 R(t;\Delta) \approx
  \begin{cases}
   \displaystyle 2 e^{-3/2} \sqrt{{6}/{\pi}} & (t \ll \tau) , \\
   \displaystyle \sqrt{{\pi \tau_{d}}/{9 t}} & (t \gg \tau) .
  \end{cases}  
\end{equation}
This gives the crossover time $\tau_{c}' \approx 0.918 \tau_{d}$. (In
the previous work, we reported $\tau_{c}' \approx \tau_{d}$
\cite{Uneyama-Akimoto-Miyaguchi-2012}. This small discrepancy may be due to the accuracy of the
fitting for the short observation time region.)
To check whether Eq.~\eqref{rf_asymptotic} actually holds or not, here we examine the RSD and RF data obtained by the discrete
reptation model.
We show the RSD and RF of the
TAMSD of a CM in the reptation model for $Z = 80$ and $\Delta = 10
\tau_{l}$ ($\tau_{l}$ is the characteristic time of the longitudinal
segmental motion) in Figure \ref{rsd_and_rf_tamsd_reptation}. We
also show the asymptotic forms of the RF calculated from the asymptotic
forms of the RSD by
Eq.~\eqref{rf_asymptotic_reptation} in Figure \ref{rsd_and_rf_tamsd_reptation}.
From Figure \ref{rsd_and_rf_tamsd_reptation}, we find that our
theoretical prediction agrees well with the asymptotic behavior of the
simulation data.
Thus we conclude that the RF shows qualitatively the same behavior as
the RSD.
Both the RSD and RF of the TAMSD can be utilized
to study the long time relaxation behavior.

\section{Detailed Calculations for Reptation Model}
\label{detailed_calculations_for_reptation_model}

In this appendix, we show the detailed calculation for the correlation
function $\psi_{1}(t)$ in the reptation model.
In the reptation model, many of dynamical quantities can be calculated
from the tube survival probability, which represents the probability of
a tube segment at time $0$ survives up to time $t$ \cite{Doi-Edwards-book}. The tube survival
probability of the segment index $s$ at time $t$ can be expressed as
\begin{equation}
 \label{surviving_function}
 \Psi(s;t) = \sum_{k : \text{odd}} \frac{4}{k \pi}
  \sin\left(\frac{k \pi s}{Z}\right) \exp\left(-\frac{k^{2}
       t}{\tau_{d}}\right) .
\end{equation}
Here $Z$ is the number of tube segments ($0 \le s \le Z$) and $\tau_{d}$
is the disengagement time. Note that the expression of the surviving
probability in this work is slightly different from commonly utilized
one in the Doi-Edwards textbook \cite{Doi-Edwards-book}.
In this work $s$ represents the segment index along
the tube ($0 \le s \le Z$) whereas in the Doi-Edwards definition $s$ represents the
distance along the tube ($0 \le s \le Z a$).
(Our definition makes the calculations
slightly simple, as shown below.)

To calculate the higher order correlation functions, we need 
the joint survival probability $\Psi(s,s';t)$ of two segment indices $s$ and
$s'$. $\Psi(s,s';t)$ represents the probability that both of segments
$s$ and $s'$ at time $0$ survive up to time $t$. $\Psi(s,s';t)$ can be
obtained by solving the first-passage type problem.
(This is in a similar way to
the calculation of $\Psi(s,t)$.) We consider the case $s \le s'$, and set $\xi \equiv s' - s$.
Then $\Psi(s,s';t) = \Psi(s,s + \xi;t)$ obeys the backward Fokker-Planck equation:
\begin{equation}
 \label{backward_fokker_planck}
 \frac{\partial \Psi(s,s + \xi;t)}{\partial t}
  = \frac{1}{Z \tau_{l}} \frac{\partial^{2} \Psi(s,s + \xi;t)}{\partial
  s^{2}} ,
\end{equation}
with $\tau_{l}$ being the characteristic time scale of the longitudinal
motion of a segment along the tube. The initial condition for
Eq.~\eqref{backward_fokker_planck} is
\begin{equation}
 \label{initial_condition}
 \Psi(s,s + \xi;0) = 1 ,
\end{equation}
and the boundary condition for Eq.~\eqref{backward_fokker_planck} is
\begin{equation}
 \label{boundary_condition}
 \Psi(Z - \xi,Z;t) = \Psi(0,\xi;t) = 0 .
\end{equation}
 The disengagement time
(the longest relaxation time) $\tau_{d}$ is related to $\tau_{l}$ as $\tau_{d} = Z^{3} \tau_{l} / \pi^{2}$.
By solving Eq.~\eqref{backward_fokker_planck}, we have the following
expression for the joint survival probability:
\begin{equation}
 \label{surviving_function_2}
 \Psi(s,s + \xi;t) = \sum_{k : \text{odd}} \frac{4}{k \pi}
  \sin\left(\frac{p \pi s}{Z - \xi}\right) \exp\left[ -\frac{Z^{2}
       k^{2} t}{(Z - \xi)^{2} \tau_{d}}\right] .
\end{equation}
For the case of $s > s'$, the solution is the same form as
Eq.~\eqref{surviving_function_2} with $s$ and $\xi$ replaced by $s'$ and
$s - s'$, respectively. Combining them and we have
\begin{equation}
 \label{surviving_function_2_final}
 \Psi(s,s';t) = \sum_{k : \text{odd}} \frac{4}{k \pi}
  \sin\left(\frac{k \pi \min(s,s')}{Z - |s - s'|}\right) \exp\left[ -\frac{Z^{2}
       k^{2} t}{(Z - |s - s'|)^{2} \tau_{d}}\right].
\end{equation}
For the case of $s = s'$, Eq.~\eqref{surviving_function_2_final} reduces
to $\Psi(s;t)$:
\begin{equation}
 \label{surviving_function_2_reduction}
 \Psi(s,s;t) = \Psi(s;t) .
\end{equation}

We consider the four-body two-time correlation function $\psi_{1}(t)$.
From Eq.~\eqref{correlation_function_psi1_reptation}, we need to
calculate the correlation function of the end-to-end vector, $\langle
\bm{p}^{2}(t) \bm{p}^{2}(0) \rangle$. It can be calculated by utilizing
the joint survival probability $\Psi(s,s';t)$,
because the end-to-end vector can
be expressed in terms of the bond vectors of segments.
The end-to-end vector at time $t$ can be expressed as
\begin{equation}
 \label{end_to_end_vector}
 \bm{p}(t) = \int_{0}^{Z} ds \, \bm{u}(s,t) .
\end{equation}
where $\bm{u}(s,t)$ is the bond vector at the segment index $s$ at time $t$.
The bond vector obeys the Gaussian statistics in equilibrium. The
first and second moments in equilibrium are given as
\begin{equation}
 \langle \bm{u}(s) \rangle = 0, \qquad
  \langle \bm{u}(s) \bm{u}(s') \rangle = \frac{1}{3} a^{2} \bm{1}
  \delta(s - s') .
\end{equation}
Here $a$ is the tube segment size.
The average end-to-end vector size can be calculated straightforwardly:
\begin{equation}
 \langle \bm{p}^{2} \rangle = Z a^{2} .
\end{equation}

The correlation function $\psi_{1}(t)$
correlation function can be evaluated if we know whether two bonds
at time $0$ still survive at time $t$ or not. There are three possible cases.
The first case is where both two bonds ($s$ and $s'$) survive,
and the probability for this case is given as $\Psi(s,s';t)$. The second
case is where only one bond ($s$ or $s'$) survives, and the probabilities
are given as $\Psi(s;t) - \Psi(s,s';t)$ or $\Psi(s';t) - \Psi(s,s';t)$,
respectively. The third case is where none of two bonds survives, and
the probability for this case is $1 - \Psi(s;t) - \Psi(s';t) + \Psi(s,s';t)$.
Thus we have
\begin{equation}
 \label{fourth_order_end_to_end_correlation_reptation}
  \begin{split}
  \langle \bm{p}^{2}(t) \bm{p}^{2}(0) \rangle
  & = \int_{0}^{Z} ds \int_{0}^{Z} ds' \int_{0}^{Z} dv \int_{0}^{Z} dv' \,
   \bigg[ \langle [\bm{u}(s) \cdot \bm{u}(s')] [\bm{u}(v) \cdot
   \bm{u}(v')] \rangle \Psi(s,s';t) \\
   & \qquad + \langle \bm{u}(s') \rangle \cdot \langle \bm{u}(s) [\bm{u}(v) \cdot
   \bm{u}(v')] \rangle (\Psi(s;t) - \Psi(s,s';t)) \\
   & \qquad + \langle \bm{u}(s) \rangle \cdot \langle \bm{u}(s') [\bm{u}(v) \cdot
   \bm{u}(v')] \rangle (\Psi(s';t) - \Psi(s,s';t)) \\
   & \qquad + \langle \bm{u}(s) \cdot \bm{u}(s') \rangle \langle \bm{u}(v) \cdot
   \bm{u}(v') \rangle (1 - \Psi(s;t) - \Psi(s';t) + \Psi(s,s';t)) \bigg] \\
  & = \int_{0}^{Z} ds \int_{0}^{Z} ds' \int_{0}^{Z} dv \int_{0}^{Z} dv' \,
   \langle [\bm{u}(s) \cdot \bm{u}(s')] [\bm{u}(v) \cdot
   \bm{u}(v')] \rangle \Psi(s,s';t) \\
   & \qquad + Z^{2} a^{4} + Z a^{4} \int_{0}^{Z} ds \,
   (\Psi(s,s;t) - 2 \Psi(s;t)) .
  \end{split}
\end{equation}
By using the Wick's theorem\cite{Doi-Edwards-book}, the average for the
bond vectors in
Eq.~\eqref{fourth_order_end_to_end_correlation_reptation} can be decomposed as follows:
\begin{equation}
 \label{fourth_order_bond_correlation_reptation}
  \begin{split}
   & \langle [\bm{u}(s) \cdot \bm{u}(s')] [\bm{u}(v) \cdot
   \bm{u}(v')] \rangle \\
   & = \langle \bm{u}(s) \cdot \bm{u}(s') \rangle \langle \bm{u}(v) \cdot \bm{u}(v')  \rangle
   + \langle \bm{u}(s) \bm{u}(v) \rangle : \langle \bm{u}(s') \bm{u}(v') \rangle 
   +  \langle \bm{u}(s) \bm{u}(v') \rangle : \langle \bm{u}(s') \bm{u}(v) \rangle \\
   & = a^{4} \delta(s - s') \delta(v - v')
   + \frac{a^{4}}{3} \delta(s - v) \delta(s' - v')
   + \frac{a^{4}}{3} \delta(s - v') \delta(s' - v) .
  \end{split}
\end{equation}
The first term in the last line of
Eq.~\eqref{fourth_order_end_to_end_correlation_reptation} is calculated to be
\begin{equation}
 \label{fourth_order_end_to_end_correlation_reptation_modified}
 \begin{split}
  & \int_{0}^{Z} ds \int_{0}^{Z} ds' \int_{0}^{Z} dv \int_{0}^{Z} dv' \,
   \langle [\bm{u}(s) \cdot \bm{u}(s')] [\bm{u}(v) \cdot
   \bm{u}(v')] \rangle \Psi(s,s';t) \\
  & =  Z a^{4} \int_{0}^{Z} ds \, \Psi(s,s;t)
   + \frac{2 a^{4}}{3} \int_{0}^{Z} ds
  \int_{0}^{Z} ds' \, \Psi(s,s';t) .
 \end{split}
\end{equation}
From Eqs.~\eqref{correlation_function_psi1_reptation},
\eqref{fourth_order_end_to_end_correlation_reptation}, and
\eqref{fourth_order_end_to_end_correlation_reptation_modified},
$\psi_{1}(t)$ can be simplified:
\begin{equation}
 \label{correlation_function_psi1_reptation_modified}
  \begin{split}
   \psi_{1}(t)
  & = \frac{2}{Z} \int_{0}^{Z} ds \,
   [\Psi(s,s;t) - \Psi(s;t)]
   + \frac{2}{3 Z^{2}} \int_{0}^{Z} ds
  \int_{0}^{Z} ds' \, \Psi(s,s';t) \\
  & = \frac{2}{3 Z^{2}} \int_{0}^{Z} ds
  \int_{0}^{Z} ds' \, \Psi(s,s';t) ,
  \end{split}
\end{equation}
where we have used Eq.~\eqref{surviving_function_2_reduction}.

Eq.~\eqref{correlation_function_psi1_reptation_modified}
can be further modified by substituting
Eq.~\eqref{surviving_function_2_final} into it:
\begin{equation}
 \label{correlation_function_psi1_reptation_modified2}
 \begin{split}
  \psi_{1}(t)
  & = \frac{8}{3 \pi Z^{2}} \sum_{k : \text{odd}} \frac{1}{k} \int_{0}^{Z} ds
  \int_{0}^{Z} ds' \, 
  \sin\left[\frac{k \pi  \min(s,s')}{Z - |s - s'|}\right] \exp\left[ -\frac{Z^{2}
       k^{2} t}{(Z - |s - s'|)^{2} \tau_{d}}\right] \\
  & = \frac{16}{3 \pi Z^{2}} \sum_{k : \text{odd}} \frac{1}{k}
  \int_{0}^{Z} ds  \int_{0}^{s} ds' \, 
  \sin\left[\frac{k \pi s'}{Z - (s - s')}\right] \exp\left[ -\frac{Z^{2}
       k^{2} t}{(Z - (s - s'))^{2} \tau_{d}}\right] .
 \end{split}
\end{equation}
By introducing the variable transform $w \equiv s - s'$,
Eq.~\eqref{correlation_function_psi1_reptation_modified2} can be
integrated over $s$ as
\begin{equation}
 \label{correlation_function_psi1_reptation_modified3}
 \begin{split}
  \psi_{1}(t) 
  & = \frac{16}{3 \pi Z^{2}}  \sum_{k : \text{odd}} \frac{1}{k} \int_{0}^{Z} dw
  \int_{w}^{Z} ds \, 
  \sin\left[\frac{k \pi (s - w)}{Z - w}\right] \exp\left[ -\frac{Z^{2}
       k^{2} t}{(Z - w)^{2} \tau_{d}}\right] \\
  & = \frac{32}{3 \pi^{2} Z^{2}} \sum_{k : \text{odd}} \frac{1}{k^{2}}
  \int_{0}^{Z} dw \,
  (Z - w) \exp\left[ -\frac{Z^{2} k^{2} t}{(Z - w)^{2} \tau_{d}}\right] .
 \end{split}
\end{equation}
We introduce another variable transform $x \equiv Z^{2} / (Z - w)^{2}$ to make
the integral simple and tractable:
\begin{equation}
 \label{correlation_function_psi1_reptation_modified4}
 \begin{split}
  \psi_{1}(t) 
  & =  \frac{16}{3 \pi^{2}} \sum_{k : \text{odd}} \frac{1}{k^{2}} \int_{1}^{\infty}
  dx \, x^{-2} \exp\left( -\frac{k^{2} t}{\tau_{d}} x \right) \\
  & =  \frac{16}{3 \pi^{2}} \sum_{k : \text{odd}} \frac{1}{k^{2}}
  E_{2}(k^{2} t / \tau_{d})  .
 \end{split}
\end{equation}
In the last line of
Eq.~\eqref{correlation_function_psi1_reptation_modified4}, we have
utilized the definition of the (generalized) exponential integral\cite{NIST-handbook}.
Thus we have the explicit expression for the correlation function
$\psi_{1}(t)$ in
the main text, Eq.~\eqref{correlation_function_psi1_reptation_final}.

{
\section{Decoupling Approximation for Reptation Model}
\label{decoupling_approximation_for_reptation_model}

In the main text, we employed the decoupling approximation for the
reptation model without any
justifications. The dynamics of the
end-to-end vector depends on the one-dimensional white noise $w(t)$, and
thus the decoupling approximation seems not to be fully justified.
Here we discuss the validity of the decoupling
approximation for the reptation model and show that the decouplilng
approximation is reasonable for our calculations.

We consider the EAMSD (Eq.~\eqref{eamsd}) for the
reptation model without the decoupling approximation:
\begin{equation}
 \label{eamsd_reptation}
  \langle [\bm{r}(\Delta) - \bm{r}(0)]^{2} \rangle
  = \frac{6 D_{\text{CM}}}{\langle \bm{p}^{2} \rangle}
  \int_{0}^{t} ds \int_{0}^{t} ds' \langle \bm{p}(s) \cdot \bm{p}(s')
  w(s) w(s') \rangle .
\end{equation}
The integrand in the right hand side of Eq.~\eqref{eamsd_reptation} can be rewritten as
\begin{equation}
 \label{eamsd_reptation_integrand}
  \langle \bm{p}(s) \cdot \bm{p}(s') w(s) w(s') \rangle
  = \left\langle \langle \bm{p}(s) \cdot \bm{p}(s') \rangle_{w(s),w(s')} w(s) w(s') \right\rangle
\end{equation}
where $\langle \dots \rangle_{w(s),w(s')}$ represents the ensemble
average under given values of $w(s)$ and $w(s')$. Without loss of generality, we
consider the case of $s \geq s'$.
The correlation function $\langle \bm{p}(s) \cdot \bm{p}(s')
\rangle_{w(s),w(s')}$ can be expressed in terms of the fraction of the surviving
tube segments at time $s'$. The fraction of surviving tube segments is
related to the minimum and maximum values of the displacement of the
chain along the tube. Therefore,
the correlation function can be calculated as follows:
\begin{equation}
 \label{eamsd_reptation_integrand_modified2}
  \langle \bm{p}(s) \cdot \bm{p}(s') \rangle_{w(s) w(s')}
  = \langle \bm{p} \rangle^{2}
  \bigg\langle \max \bigg[ 0, Z
  + \min_{\substack{s'' \\ (s' < s'' < s)}} W(s'',s')
  - \max_{\substack{s'' \\ (s' < s'' < s)}} W(s'',s') \bigg]
  \bigg\rangle_{w(s),w(s')}
\end{equation}
Here $W(s,s')$ is the one-dimensional displacement of the chain along
the tube, at time $s$ starting from time $s'$. $W(s,s')$ can be
expressed as a time integral of the one-dimensional noise $w(t)$:
\begin{equation}
 \label{one_dimensional_displacement}
 W(s,s') \equiv \sqrt{\frac{2}{Z \tau_{l}}} \int_{s'}^{s} du \, w(u) .
\end{equation}
The contributions of $w(s)$ and $w(s')$ to $W(s,s')$ are infinitesimally
small, and thus from Eq.~\eqref{eamsd_reptation_integrand_modified2},
$\langle \bm{p}(s) \cdot \bm{p}(s') \rangle_{w(s) w(s')}$ becomes
statistically independent of $w(s)$ and $w(s')$. Finally
Eq.~\eqref{eamsd_reptation_integrand} becomes
\begin{equation}
 \label{eamsd_reptation_integrand_final}
  \langle \bm{p}(s) \cdot \bm{p}(s') w(s) w(s') \rangle
  = \langle \bm{p}(s) \cdot \bm{p}(s') \rangle \langle w(s) w(s') \rangle
\end{equation}
From the symmetry, Eq.~\eqref{eamsd_reptation_integrand_final} also
holds for $s < s'$.
Eq.~\eqref{eamsd_reptation_integrand_final} justifies the decoupling
approximation for the calculation of the EAMSD.

For the calculation of the RSD of the TAMSD, we have a similar
correlation function (see Eq.~\eqref{tamsd_square_average_initial} in Appendix~\ref{detailed_calculations_for_ensemble_average_of_squared_time-averaged_mean_square_displacement}):
\begin{equation}
 \label{rsd_reptation_integrand}
  \begin{split}
   & \langle \bm{p}(s_{1}) \cdot \bm{p}(s_{2}) \bm{p}(s_{3}) \cdot
  \bm{p}(s_{4}) w(s_{1}) w(s_{2}) w(s_{3}) w(s_{4}) \rangle \\
   & = \left\langle \langle \bm{p}(s_{1}) \cdot \bm{p}(s_{2}) \bm{p}(s_{3}) \cdot
  \bm{p}(s_{4}) \rangle_{w(s_{1}),w(s_{2}),w(s_{3}),w(s_{4})} w(s_{1})
   w(s_{2}) w(s_{3}) w(s_{4}) \right\rangle .
  \end{split}
\end{equation}
Here $\langle \dots \rangle_{w(s_{1}),w(s_{2}),w(s_{3}),w(s_{4})}$
represents the ensemble average under given
$w(s_{1}),w(s_{2}),w(s_{3})$, and $w(s_{4})$. 
The correlation function $\langle \bm{p}(s_{1}) \cdot \bm{p}(s_{2}) \bm{p}(s_{3}) \cdot
  \bm{p}(s_{4}) \rangle_{w(s_{1}),w(s_{2}),w(s_{3}),w(s_{4})}$ depends
  only on the one-dimensional displacement along the
tube. (The explicit
expression becomes quite complicated.) The contributions of $w(s_{1}),w(s_{2}),w(s_{3})$, and
$w(s_{4})$ to the one-dimensional displacement are infinitesimally small, as
the previous case. Therefore we find that
Eq.~\eqref{rsd_reptation_integrand} can be rewritten as the following decoupled form:
\begin{equation}
 \label{rsd_reptation_integrand_final}
  \begin{split}
   & \langle \bm{p}(s_{1}) \cdot \bm{p}(s_{2}) \bm{p}(s_{3}) \cdot
  \bm{p}(s_{4}) w(s_{1}) w(s_{2}) w(s_{3}) w(s_{4}) \rangle \\
   & =\langle \bm{p}(s_{1}) \cdot \bm{p}(s_{2}) \bm{p}(s_{3}) \cdot
  \bm{p}(s_{4}) \rangle \langle w(s_{1}) w(s_{2}) w(s_{3}) w(s_{4})
   \rangle .
  \end{split}
\end{equation}
This justifies the use of the decoupling approximation for the
calculation of the RSD of the TAMSD.
}

\section{Detailed Calculations for Non-Markovian Two-State Model}
\label{detailed_calculations_for_non_markovian_two_state_model}

In this appendix, we show the detailed calculations for the RSD of the TAMSD in the non-Markovian two-state model.
The (unilateral) Laplace transform is convenient to
calculate some quantities for such a non-Markovian model.
For example, the Laplace
transform of the equilibrium trapping-time distribution
$\rho_{h}^{\text{(eq)}}(t)$
(Eq.~\eqref{trapping_time_distribution_equilibrium_explicit}) simply becomes
\begin{equation}
  \hat{\rho}_h^{\mathrm{(eq)}} (u) = \frac {1 - \hat{\rho}_h(u)}{u \langle \tau \rangle_{h}}. 
\end{equation}
Here, the functions with hats (such as $\hat{\rho_{h}}$ and
$\hat{\rho}_{h}^{\text{(eq)}}$) represent the Laplace transformed functions.
We define the distribution for the
sum of two successive trapping times as $\rho(\tau)$. $\rho(\tau)$ can
be expressed as the convolution of $\rho_{f}(\tau)$ and $\rho_{s}(\tau)$,
\begin{equation}
 \rho(\tau) \equiv \rho_{f} * \rho_{s}(\tau) = \rho_{s} * \rho_{f} (\tau) = \int_{0}^{\tau} d\tau' \, \rho_{f}(\tau - \tau') \rho_{s}(\tau') .
\end{equation}
The Laplace transform of $\rho(\tau)$ simply becomes
\begin{equation}
  \hat{\rho}(u) = \hat{\rho}_f(u) \hat{\rho}_{s}(u) .
\end{equation}

We express the probabilities of
having $n$ transitions up to time $t$ starting from state $h$ at time $0$, as
$Q_{h,n}(t)$ $(h = f,s)$.
For convenience, we introduce the following integral operator $\mathcal{I}$:
\begin{equation}
 \mathcal{I} f(t) \equiv \int_{t}^{\infty} dt' f(t') .
\end{equation}
Then, $Q_{h,n}(t)$ can be expressed as \cite{Godreche-Luck-2001}
\begin{equation}
 Q_{h,n}(t) =
  \begin{cases}
   \mathcal{I} \rho_{h}^{\text{(eq)}}(t) & (n = 0), \\
   \rho_{h}^{\text{(eq)}} * \overbrace{\rho * \rho * \dots * \rho}^{(n - 1) / 2} * (\mathcal{I} \rho_{\bar{h}}) (t) &
   (n = 1,3,5,\dots) , \\
   \rho_{h}^{\text{(eq)}} * \overbrace{\rho * \rho * \dots *
   \rho}^{n / 2 - 1} * \, \rho_{\bar{h}} * (\mathcal{I} \rho_{h}) (t) & (n
   = 2,4,6,\dots) .
  \end{cases}
\end{equation}
where $\bar{h} = s$ and $h$ for $h=f$ and $s$, respectively.
The Laplace transform of $Q_{h,n}(t)$ becomes
\begin{equation}
 \label{jump_number_distribution_nonmarkov_laplace_transform}
  \hat{Q}_{h,n}(u) =
  \begin{cases}
    \displaystyle
    \frac {\langle \tau
   \rangle_{h} u - 1 + \hat{\rho}_{h}(u)}{\langle \tau
   \rangle_{h} u^{2}}
    &(n=0), \\
   \displaystyle
   \frac{[1 - \hat{\rho}_{h}(u)] [1 - \hat{\rho}_{\bar{h}}(u)]}{\langle \tau \rangle_{h} u^{2}}
   \hat{\rho}^{(n - 1) / 2}(u)
    & (n=1,3,5,\dots), \\
    \displaystyle
   \frac{[1 - \hat{\rho}_h(u)]^{2}}{\langle \tau \rangle_{h} u^{2}}
   \hat{\rho}_{\bar{h}}(u) \hat{\rho}^{n / 2 -1}(u)
    & (n = 2, 4, 6, \dots) .
  \end{cases}
\end{equation}
The transition probability $W_{hh'}(t)$ can be expressed in terms of
$Q_{h,n}(t)$ as
\begin{equation}
 \label{transition_probability_nonmarkov}
  W_{hh'}(t) =
   \begin{cases}
    \displaystyle \sum_{n=0}^{\infty} Q_{h,2n} (t) & (h' = h),  \\
        \displaystyle \sum_{n=0}^{\infty} Q_{\bar{h},2n+1}(t) & (h' = \bar{h}) .
   \end{cases}
\end{equation}

We calculate the asymptotic form of the correlation function
$\psi_{1}(t)$ in the long time region. The long time asymptotic behavior
can be calculated from the small $u$ limit for the Laplace transform.
Since $W_{hh'}(t)$ converges to $\phi_h$ at the limit of $t\to \infty$, it is
convenient to consider $W_{hh'}(t) - \phi_{h}$ rather than $W_{hh'}(t)$
itself.
From Eqs.~\eqref{jump_number_distribution_nonmarkov_laplace_transform}
and \eqref{transition_probability_nonmarkov}, we obtain
the following asymptotic form for the Laplace transform of $W_{hh'}(t) - \phi_{h}$
for small $u$:
\begin{equation}
 \label{transition_probability_nonmarkov_asymptotic}
  \hat{W}_{hh'}(u) - \frac {\phi_{h}}{u} \approx
   \sigma_{h h'} \frac{\langle \tau^{2} \rangle_{s} \langle \tau \rangle_{f}^{2}
   + \langle \tau^{2} \rangle_{f} \langle \tau \rangle_{s}^{2}
   - 2 \langle \tau \rangle_{s}^{2} \langle \tau \rangle_{f}^{2}}
   {2 \langle \tau \rangle_{h'} (\langle \tau \rangle_{s} +
   \langle \tau \rangle_{f})^{2}} ,
\end{equation}
where $\sigma_{h h'} = 1$ or $-1$ for $h' = h$ or $h' = \bar{h}$,
respectively, and we have utilized the expansion of $\hat{\rho}_{h}(u)$
around $u = 0$,
\begin{equation}
\hat{\rho}_{h}(u) = 1 - \langle \tau \rangle_{h} u - \langle
  \tau^{2} \rangle_{h} u^{2} / 2 + \dotsb.
\end{equation}
From Eq.~\eqref{transition_probability_nonmarkov_asymptotic},
we have the following simple relation for the transition probability:
\begin{equation}
 \label{integral_transition_probability_asymptotic}
 \int_{0}^{\infty} dt \, [W_{hh'}(t) - \phi_h]=
  \lim_{u \to 0} \left[\hat{W}_{hh'}(u) - \frac {\phi_{h}}{u} \right]
  =
  \begin{cases}
   \tilde{\tau} \phi_{\bar{h}} & (h' = h), \\
   - \tilde{\tau} \phi_{h} & (h' = \bar{h}) .
  \end{cases}
\end{equation}
where $\tilde{\tau}$ is the characteristic relaxation time defined by Eq.~\eqref{relaxation_time_nonmarkov}.
By combining Eqs.~\eqref{psi1_two_state_nonmarkov} and \eqref{integral_transition_probability_asymptotic},
finally we have Eq.~\eqref{psi1_integral_nonmarkov}.

\section{Non-Gaussianity Parameter}
\label{non_gaussianity_parameter}

The non-Gaussianity
parameter \cite{Rahman-1964,Ernst-Kohler-Weiss-2014,Cherstvy-Metzler-2014}
is widely employed to
investigate the non-Gaussian properties of the diffusion processes.
In this appendix, we calculate the expression for the non-Gaussian
parameter $A(\Delta)$ (Eq.~\eqref{non_gaussianity_parameter_definition})
in terms of the four-body two-time correlation functions. Then we compare it with the RSD of the TAMSD.

The ensemble average of quartic displacement can be calculated in the
same way as Eq.~\eqref{tamsd_square_average},
\begin{equation}
 \label{ensemble_average_quartic_displacement}
  \begin{split}
   \langle [ \bm{r}(\Delta) - \bm{r}(0) ]^{4} \rangle 
   & = 4 \int_{0}^{\Delta} ds \int_{0}^{\Delta} ds'
     \int_{0}^{\Delta} du \int_{0}^{\Delta} du' \, \langle w_{i}(s) w_{j}(s') w_{k}(u) w_{l}(u') \rangle\\
   & \qquad \times
   \langle B_{mi}(s) B_{mj}(s') B_{nk}(u) B_{nl}(u') \rangle \\
   & = 8 \int_{0}^{\Delta} ds 
     \int_{0}^{s} du \, \langle \tr \bm{D}(s) \tr \bm{D}(u) \rangle \\
   & \qquad + 16 \int_{0}^{\Delta} ds \int_{0}^{s} du \,
   \tr \langle \bm{D}(s) \cdot \bm{D}(u) \rangle .
  \end{split}
\end{equation}
By using the correlation functions $\psi_{1}(t)$ and $\psi_{2}(t)$
defined in Eqs.~\eqref{correlation_fucntion_psi1_definition} and 
\eqref{correlation_fucntion_psi2_definition}, Eq.~\eqref{ensemble_average_quartic_displacement} can be rewritten as
\begin{equation}
 \label{ensemble_average_quartic_displacement_modified}
  \begin{split}
   \langle [ \bm{r}(\Delta) - \bm{r}(0) ]^{4} \rangle 
   & = 4 \left(1 + \frac{2 C}{n} \right)
   [\tr \langle \bm{D} \rangle]^{2} \Delta^{2} \\
   & \qquad + 8 [\tr \langle \bm{D} \rangle]^{2} \int_{0}^{\Delta} ds 
     \int_{0}^{s} du \, [ \psi_{1}(s - u) +
   2 C \psi_{2}(s - u)] . \\
  \end{split}
\end{equation}
From Eqs.~\eqref{non_gaussianity_parameter_definition} and
\eqref{ensemble_average_quartic_displacement_modified}, finally we have
the following formula for the non-Gaussianity parameter:
\begin{equation}
 \label{non_gaussianity_paramter_formula}
  A(\Delta)
   = \frac{2 (C - 1)}{n + 2} + \frac{2 n}{(n + 2)
   \Delta^{2}} \int_{0}^{\Delta} ds 
     \int_{0}^{s} du \, [ \psi_{1}(s - u) +
   2 C \psi_{2}(s - u)] .
\end{equation}

Eq.~\eqref{non_gaussianity_paramter_formula} contains both
$\psi_{1}(t)$ and $\psi_{2}(t)$. Because these correlation functions
exhibit the characteristic long time relaxation, the non-Gaussianity parameter can be
utilized to analyze the characteristic relaxation at the long time
scale. The short and long time asymptotic forms are calculated to be
\begin{equation}
 \label{non_gaussianity_parameter_asymptotic}
  A(\Delta) \approx
  \begin{cases}
   \displaystyle \frac{2 (C - 1)}{n + 2} + \frac{n}{n + 2}[ \psi_{1}(0) +
   2 C \psi_{2}(0)] 
   & (\Delta \ll \tau) , \\
   \displaystyle \frac{2 (C - 1)}{n + 2} + \frac{2 n}{(n + 2) \Delta}
     \int_{0}^{\infty} dv \, [ \psi_{1}(v) +
   2 C \psi_{2}(v)]
   & (\Delta \gg \tau)  .
  \end{cases}
\end{equation}
Eq.~\eqref{non_gaussianity_parameter_asymptotic} has somewhat similar
properties to the squared RSD, Eq.~\eqref{rsd_square_asymptotic}.
However, the bahavior of $A(\Delta)$ is qualitatively different from one
of the squared RSD. $A(\Delta)$ approaches to the constant $2
(C - 1)  / (n + 2)$ at the limit of $\Delta \to \infty$, whereas $\Sigma^{2}(t;\Delta)$ approaches to zero at
the limit of $t \to \infty$. Such a property of $A(\Delta)$ makes the
numerical analysis difficult. (We need to determine the constant $2
(C - 1)  / (n + 2)$ and then subtract it from $A(\Delta)$.) In the case
of the isotropic systems, $C = 1$ and this constant vanishes. Then Eq.~\eqref{non_gaussianity_parameter_asymptotic}
reduces to
\begin{equation}
 \label{non_gaussianity_parameter_asymptotic_isotropic}
  A(\Delta) \approx
  \begin{cases}
   \displaystyle \frac{n}{n + 2}[ \psi_{1}(0) +
   2 \psi_{2}(0)] 
   & (\Delta \ll \tau) , \\
   \displaystyle \frac{2 n}{(n + 2) \Delta}
     \int_{0}^{\infty} dv \, [ \psi_{1}(v) +
   2 \psi_{2}(v)]
   & (\Delta \gg \tau)  .
  \end{cases}
\end{equation}
Even in this simple case, $A(\Delta)$ depends both on $\psi_{1}(t)$ and
$\psi_{2}(t)$. Also, $A(\Delta)$ explicitly depends on the dimension of
the system.
On the other hand, the explicit expression for the squared RSD
(Eq.~\eqref{rsd_square_final}) and its asymptotic forms
(Eq.~\eqref{rsd_square_asymptotic}) are simple and common for isotropic and
anisotropic systems. (As mentioned in the main text, $\Sigma^{2}(t;\Delta)$ essentially depends only on
$\psi_{1}(t)$.) Thus we
consider that the RSD would be more suitable than the non-Gaussianity
parameter, to characterize the long time relaxation behavior of
time-dependent and fluctuating diffusivities.

\bibliographystyle{apsrev4-1}
\bibliography{time_dependent_diffusivity}


\clearpage

\section* {Figure Captions}

\hspace{-\parindent}%
Figure \ref{rsd_tamsd_reptation}:
The RSD of the TAMSD of a CM in the discrete reptation model.
The number of tube segments $Z$ is $Z = 80$ and the time difference $\Delta$ is $\Delta = 10 \tau_{l}$.
$\tau_{l}$ is the characteristic time of the longitudinal motion of a segment along the tube.
Symbols represent the kinetic Monte Carlo simulation data. The dotted
and dashed
curves represent the theoretical prediction (Eq.~\eqref{rsd_square_explicit_reptation}) and its
asymptotic forms (Eq.~\eqref{rsd_asymptotic_reptation}).

\

\hspace{-\parindent}%
Figure \ref{rsd_tamsd_two_state_model}:
The RSD of the TAMSD of the Markovian two-state model. The diffusion
coefficients and transition
rates are  $D_{s} = 1$, $D_{f} = 10$, $k_{s} = 1$ and
$k_{f} = 0.1, 1,$ and $10$. The time difference is $\Delta =
0.001$. Symbols represent the simulation results
and solid curves represent the theoretical prediction.

\

\hspace{-\parindent}%
Figure \ref{rsd_tamsd_two_state_model_nonmarkov}:
The RSD of the TAMSD of the non-Markovian two-state model. The
diffusion coefficients are $D_{s} = 1$ and $D_{f} = 10$. The waiting
time distribution for the fast state is given by the exponential
distribution with $k_{f} = 1$, and the waiting time distribution for the
slow state is given by the power-law type distribution with $k_{s,1} = 1$
and $k_{s,0} = 0.1, 0.01,$ and $0.001$. The time difference is $\Delta =
0.001$.
Symbols represent the simulation results
and dashed curves represent the theoretically predicted asymptotic forms.

\

\hspace{-\parindent}%
Figure \ref{rsd_and_rf_tamsd_reptation}:
The RSD and RF of the TAMSD of a CM in the discrete reptation model.
(The RF data are taken from Ref \cite{Uneyama-Akimoto-Miyaguchi-2012}.)
The number of tube segments $Z = 80$ and the time difference $\Delta =
10 \tau_{l}$, where $\tau_{l}$ is the characteristic time of the
longitudinal motion of a segment along the tube.
Dashed curves represent the asymptotic forms 
for the RF  
(Eq.~\eqref{rf_asymptotic_reptation}).

\clearpage

\section* {Figures}

\clearpage

\clearpage

\clearpage

\clearpage

\end{document}